\documentclass[hyper,11pt,paper]{article}
\usepackage{jheppub} 

\usepackage[T1]{fontenc} 
\usepackage{graphicx,amsmath,amssymb,multirow,array,bm,bbm,mathrsfs}
\usepackage[mathscr]{eucal}

\newcommand{\GN}{G_{_N}}

\newcommand{\lAdS}{\ell_\text{AdS}}
\newcommand{\lAdSt}{\tilde{\ell}_\text{AdS}}
\newcommand{\lP}{\ell_\text{P}}
\newcommand{\lPt}{{{\ell}}_\text{P}}

\newcommand{\ellt}{\tilde{\ell}}
\newcommand{\gn}{\bar g}

\newcommand{\Rn}{\bar R}
\newcommand{\Ro}{{}^{\text{\tiny (1)}}\!R}
\newcommand{\gnt}{\gn}

\newcommand{\Gamman}{\bar{\Gamma}}
\newcommand{\Gammao}{{}^{\text{\tiny (1)}}{\Gamma}}

\newcommand{\In}{{I}^{\text{\tiny (0)}}}
\newcommand{\Sn}{{S}^{\text{\tiny (0)}}}
\newcommand{\Hn}{{H}_\mathcal{A}^{\text{\tiny (0)}}}

\newcommand{\dn}{\bar{\nabla}}
\newcommand{\Boxn}{\bar{\Box}}

\newcommand{\order}{\mathscr{O}}
\newcommand{\lphi}{\lambda_\phi}
\newcommand{\lphit}{\tilde\lambda_\phi}
\newcommand{\lh}{\lambda_h}
\newcommand{\lht}{\tilde{\lambda}_h}

\newcommand{\cOO}{C_{\mathcal{O}}}
\newcommand{\prop}{\mathcal{G}}

\newcommand{\CT}{C_{_T}}

\title{Comments on universal properties of entanglement entropy and bulk reconstruction}

\author[]{Felix M. Haehl}
\affiliation[]{Centre for Particle Theory \& Department of Mathematical Sciences,\\
                     Science Laboratories, South Road, Durham DH1 3LE, UK.}
\emailAdd{f.m.haehl@gmail.com}

\abstract{Entanglement entropy of holographic CFTs is expected to play a crucial role in the reconstruction of semiclassical bulk gravity. We consider the entanglement entropy of spherical regions of vacuum, which is known to contain universal contributions. After perturbing the CFT with a relevant scalar operator, also the first order change of this quantity gives a universal term which only depends on a discrete set of basic CFT parameters. We show that in gravity this statement corresponds to the uniqueness of the ghost-free graviton propagator on an AdS background geometry. While the gravitational dynamics in this context contains little information about the structure of the bulk theory, there is a discrete set of dimensionless parameters of the theory which determines the entanglement entropy. We argue that for every (not necessarily holographic) CFT, any reasonable gravity model can be used to compute this particular entanglement entropy. We elucidate how this statement is consistent with AdS/CFT and also give various generalizations. On the one hand this illustrates the remarkable usefulness of geometric concepts for understanding entanglement in general CFTs. On the other hand, it provides hints as to what entanglement data can be expected to provide enough information to distinguish, e.g., bulk theories with different higher curvature couplings.}

\begin{document} 
	
\maketitle
\flushbottom

\author{}

\section{Introduction}
\label{sec:introduction}

It is by now well known that some conformal field theories (CFTs) admit a dual description in terms of semiclassical gravity in one higher dimension \cite{Aharony:1999ti}. But it is so far not clear what are the precise CFT data that one needs to know in order to reconstruct various specific features of the bulk theory. A full knowledge of the CFT 
should be exactly equivalent to full knowledge of bulk quantum gravity, the challenge being to find the precise dictionary. We wish to focus on the question which subset of CFT parameters one needs to have access to in order to reconstruct just a classical bulk geometry and its linearized dynamics. Once this question is answered in terms of fundamental CFT parameters, the practical question is which CFT observables are sensitive to these parameters in a useful way.

One hint is the fact that a local bulk metric should be represented in the boundary CFT in a non-local fashion. An important CFT feature which does have non-local properties is the entangled structure of the underlying quantum state. It has therefore been proposed that knowing enough about the entanglement structure of the CFT state is sufficient to determine the gravitational dynamics of the bulk. In particular, entanglement entropy of spatial regions in the CFT appears to be an illuminating quantity to study in this context, c.f., \cite{VanRaamsdonk:2009ar,VanRaamsdonk:2010pw,Lashkari:2013koa,Blanco:2013joa,Faulkner:2013ica}. The basic rationale for this idea is that entanglement entropy of a given spatial region $\mathcal{A}$ can be computed in semiclassical holography as the area of a certain extremal surface \cite{Ryu:2006bv,Ryu:2006ef,Hubeny:2007xt} which, of course, encodes information about the metric. By varying this entanglement entropy (for instance by changing $\mathcal{A}$, or by varying the underlying quantum state, or by deforming the CFT spectrum), the corresponding change of the minimal surface encodes information about the dynamics of the bulk metric. Making this idea precise and general would be significant progress towards our understanding of the holographic nature of quantum states and gravity. 

We wish to carefully distinguish differently strong versions of the  expectation that entanglement encodes bulk geometry. The weakest version of the {\it ``entanglement is geometry''} statement simply says that the most efficient way of calculating entanglement entropy is by using geometric methods. For holographic CFTs the paradigmatic manifestation of this statement is the Ryu-Takayanagi conjecture \cite{Ryu:2006bv,Ryu:2006ef}. But even for non-holographic CFTs there is evidence that geometric concepts sometimes provide the most natural and efficient way of computing entanglement entropy. The first such statement is the realization that vacuum entanglement entropy of spherical regions can be conformally mapped to thermal entropy which sometimes has a natural interpretation in terms of black hole thermodynamics \cite{Casini:2011kv}. Refining this idea, Faulkner \cite{Faulkner:2014jva} revisited the problem of computing the response of vacuum entanglement entropy of spheres to a deformation of the CFT by a relevant scalar operator  \cite{deHaro:2000xn,Hung:2011ta,Liu:2012eea,Nishioka:2014kpa,Rosenhaus:2014woa,Rosenhaus:2014zza}. For this computation, nothing is assumed about the existence or the nature of a holographic dual of the CFT. As a CFT calculation this is a formidable problem. Intriguingly, the most convenient way of repackaging this calculation turns out to be in terms of an auxiliary gravitational problem, where the entanglement entropy is represented as a minimal surface area in cosmological Einstein gravity responding to a scalar field perturbation. This is a powerful result, in particular in face of the fact that entanglement entropy is notoriously hard to compute otherwise. One of the goals of this paper is to elucidate from a gravitational point of view how this universality comes about and to reconcile these statements with AdS/CFT duality (where bulk descriptions generically have a structure which is not necessarily given by just Einstein gravity).

A stronger form of the ``entanglement is geometry'' conjecture is the idea that in holographic CFTs the detailed bulk dynamics can be extracted entirely from knowledge of entanglement entropy for a suitable set of regions $\mathcal{A}$. Even in the semiclassical regime, this is  a strong assertion because the detailed bulk dynamics may involve not only Einstein gravity, but also higher curvature corrections \cite{Fursaev:2013fta,Hung:2011xb,deBoer:2011wk,Dong:2013qoa,Camps:2013zua,Bhattacharyya:2013gra,Bhattacharyya:2014yga}.  The imprint of higher curvature corrections on equations of motion should then be reconstructible from CFT entanglement entropies. See \cite{Lashkari:2013koa,Faulkner:2013ica,Swingle:2014uza} for recent progress towards deriving linearized Einstein equations from a first law of entanglement \cite{Bhattacharya:2012mi,Blanco:2013joa,Wong:2013gua,Allahbakhshi:2013rda}. It will be a topic of this paper to illuminate the way in which CFT entanglement does or does not discriminate between different higher curvature interactions in the bulk. 

Let us compare these two versions of the ``entanglement is geometry'' statement. If spherical region entanglement entropy in deformations of any arbitrary CFT can be computed using linearized Einstein gravity \cite{Faulkner:2014jva}, it follows that this quantity is certainly not sufficient to distinguish even CFTs with a semiclassical gravity dual from those without; let alone distinguish a bulk theory governed by Einstein gravity from any other higher derivative theory with the same spectrum of low energy excitations. Therefore, for computing this particular type of entanglement entropy, geometry is always an excellent tool; but in order to make sure that this geometry is actually the lowest order manifestation of a semiclassical AdS/CFT duality, more input is required. Of course, the bulk reconstruction from CFT data proceeds gradually: the more CFT parameters the entanglement entropies considered are sensitive to, the more details of the bulk dynamics are a-priori expected to be derivable from it.

In this note we wish to investigate how to find entanglement entropies which one can expect to be sufficient to reconstruct various bulk features such as linearized dynamics, higher curvature couplings etc. In the context of spherical region entanglement entropy and scalar operator deformations, we will give a detailed analysis both from a CFT point of view and from the bulk perspective. 
As we will see, at the first relevant order in the perturbation, the entanglement entropy is only sensitive to few correlation functions which have a universal form (up to normalization) thanks to conformal invariance. From the bulk perspective the analogous statement is that the backreaction of a maximally symmetric geometry to the presence of a scalar probe is universal for any acceptable theory of gravity.\footnote{ By an ``acceptable theory of gravity'' we mean any semiclassical gravitational theory with the correct spectrum of excitations (in particular without ghosts). See however \S\ref{app:generalization}.} A careful comparison reveals that this entanglement data is not sufficient to conclude anything about the presence of higher curvature terms in the bulk Lagrangian. For the purpose of bulk reconstruction we will outline how to gain sensitivity to such features, but will also demonstrate explicitly that the existence of some higher curvature interactions is much easier to conclude than the precise structure of the latter.

The structure of this note is as follows. In \S\ref{sec:universality} we first review the context of our discussion and then give CFT and gravity arguments to demonstrate universal properties of entanglement entropy of spherically symmetric regions. In \S\ref{sec:exampleGB} we illustrate our abstract discussion with the example of curvature squared theories of gravity and Gauss-Bonnet theory in particular. Finally, \S\ref{sec:discussion} contains a discussion and relates to recent developments in this context. We generalize our discussion to a wider class of higher curvature theories in \S\ref{app:generalization}, allowing also for fourth order equations of motion. In \S\ref{app:Wald} we perform the most general background field expansion for gravity at second order in the metric perturbation, thus uncovering the universality of Gauss-Bonnet theory within this class of theories. 
Some facts about the first law of entanglement are reviewed in \S\ref{sec:1stLaw}. Various useful formulae can be found in \S\ref{sec:perturbation}.\newpage

\section{Universality of entanglement entropy of spherical regions}
\label{sec:universality}

In this section we analyze universal features both of CFTs and of gravitational theories, which explain universal properties of entanglement entropy of ball shaped spatial regions in CFTs deformed by a scalar operator. We will then compare these using holography and draw conclusions for bulk reconstruction from entanglement.

\subsection{Setup}
\label{sec:setup}

Consider a holographic CFT${}_d$ in the vacuum state and on Minkowski spacetime. Assume further that the theory has a classical gravity dual described by Einstein theory and consider a ball shaped spatial region $\mathcal{A}$ of radius $R$. The reduced density matrix associated with $\mathcal{A}$ is obtained by tracing out the degrees of freedom outside of $\mathcal{A}$:
\begin{equation}\label{eq:rhoDef}
\rho_\mathcal{A} = \text{Tr}_{\mathcal{A}^c}\, |0\rangle\langle 0|  \,.
\end{equation}
Without loss of generality, we will assume $\rho_{\mathcal{A}}$ is normalized, i.e., $\text{Tr} \,\rho_{\mathcal{A}} = 1$. 
The entanglement entropy of $\mathcal{A}$ is defined as the von Neumann entropy of $\rho_\mathcal{A}$ and can be computed in the bulk as the area of the minimal surface ending on the boundary of the ball \cite{Ryu:2006bv}:
\begin{equation}
\Sn_{EE}(\mathcal{A}) = -\text{Tr}( \rho_\mathcal{A} \, \log \rho_\mathcal{A}) = \frac{\text{Vol}(\mathcal{E}_{min})}{4 \GN}  \,,
\end{equation}
where $\mathcal{E}_{min}$ is the corresponding bulk minimal surface and $\GN$ denotes Newton's constant in $d+1$ dimensions.  Since the region $\mathcal{A}$ is spherical, its causal development can be mapped by a conformal transformation to a direct product of time with a maximally symmetric hyperbolic space \cite{Casini:2011kv}. This transformation further maps the CFT vacuum state to a thermal state. The extremal surface is no longer anchored at the boundary but instead wraps the horizon of a hyperbolic black hole. The computation of entanglement entropy of $\mathcal{A}$ thus reduces to a calculation of the horizon entropy of a hyperbolic black hole at temperature $T = 1/(2\pi R)$:
\begin{equation} \label{eq:EinsteinEntropy}
\Sn_{EE}(\mathcal{A}) = S_{thermal}(\mathbb{H}_{d-1}) = \frac{\text{Vol}(\mathbb{H}_{d-1})}{4\GN}\,,
\end{equation}
where $\mathbb{H}_{d-1}$ is the hyperbolic horizon slice. For notational simplicity, we will from now on work with Planck units, i.e., $\lP^{d-1} = 8\pi \GN$.

Now let us ask what changes if the bulk gravitational theory is not just Einstein gravity, but also contains some higher curvature corrections. For concreteness, consider a $d+1$-dimensional bulk theory of the form 
\begin{equation} \label{eq:HigherDerDef}
\begin{split}
  I = \frac{1}{2\lP^{d-1}} \int  \sqrt{-g} \;  \mathcal{L} \equiv \frac{1}{2\lP^{d-1}} \int  \sqrt{-g} \, \left(R + \frac{d(d-1)}{\ell^2}+ \mathcal{L}_{h.d.} \right) \,,
\end{split}
\end{equation}
where $\mathcal{L}_{h.d.}\equiv \mathcal{L}_{h.d.}(g_{ab},R_{abcd},\nabla_e R_{abcd},\ldots)$ encodes higher curvature corrections. The cosmological length scale $\ell$ is related to the AdS scale $\lAdS$; e.g., in Einstein gravity, $\ell=\lAdS$. 
The thermal entropy (and thus the entanglement entropy) is now given by the Wald entropy \cite{Iyer:1994ys} of the hyperbolic horizon \cite{Myers:2010tj,Myers:2010xs,Hung:2011xb}:
\begin{equation}\label{eq:WaldEntropy}
\begin{split}
\Sn_{EE}(\mathcal{A}) &=  S_{_{Wald}}(\mathbb{H}_{d-1}) = \frac{-2\pi}{2\lP^{d-1}} \, \int_{\mathbb{H}_{d-1}}  \sqrt{\gamma} \; \left(-2+\frac{\delta \mathcal{L}_{h.d.}}{\delta R_{abcd}} \, n_{ab}\, n_{cd}\right) \\
&= \frac{2\pi}{\pi^{d/2}} \Gamma\left(d/2\right) \, a_d^* \; \frac{\text{Vol}(\mathbb{H}_{d-1})}{\lAdS^{d-1}} \,,
\end{split}
\end{equation}
where $n_{ab}$ denotes the binormal to the hyperbolic horizon $\mathbb{H}_{d-1}$ normalized as $n^{ab}n_{ab} = -2$ and $\gamma$ is the induced metric. The constant $a_d^*$ depends on the details of the gravitational theory. For example, in Einstein gravity $\mathcal{L}_{h.d.}=0$ and we find by comparison with \eqref{eq:EinsteinEntropy}  
\begin{equation}\label{eq:adEH}
 (a_d^*)_{Einstein} = \frac{\pi^{d/2}}{\Gamma(d/2)} \frac{\lAdS^{d-1}}{\lP^{d-1}}\,.
\end{equation}
 In fact, since the Wald entropy density is constant on maximally symmetric AdS${}_{d+1}$, we can immediately read off the general expression for the constant $a_d^*$:
\begin{equation}\label{eq:adHD}
 a_d^* = \frac{\pi^{d/2}}{\Gamma(d/2)} \frac{\lAdS^{d-1}}{\lP^{d-1}} \left(1- \frac{1}{2}\frac{\delta \mathcal{L}_{h.d.}}{\delta R_{abcd}} \, n_{ab}\, n_{cd}\bigg{|}_{AdS}\right) = -\frac{\pi^{d/2}\lAdS^{d+1}}{d\, \Gamma(d/2)} \, \frac{1}{2\lP^{d-1}} \, \mathcal{L}|_{AdS}\,.
\end{equation}
Here, the second step is a simple consequence of the existence of an AdS${}_{d+1}$ background solution and the maximal symmetry of the geometry (see, e.g., \cite{Myers:2010tj} for a detailed derivation).\footnote{ The first term in the bracket of \eqref{eq:adHD} comes from the Einstein contribution in the action \eqref{eq:HigherDerDef}. In a theory $\mathcal{L} = d(d-1)/\ell^2 + \mathcal{L}_{h.d.}$ without pure Ricci scalar, this term would disappear.}
The important feature of \eqref{eq:WaldEntropy} is that the entanglement entropy is always proportional to the horizon area $\text{Vol}(\mathbb{H}_{d-1})$ (or equivalently to $\text{Vol}(\mathcal{E}_{min})$), irrespective of the details of the gravitational theory. In particular, just knowing $\Sn_{EE}(\mathcal{A})$ does not give any insight as to what higher derivative corrections the bulk dual of the given CFT may entail: any higher derivative corrections to Einstein gravity will only change the overall normalization but not the volume scaling in \eqref{eq:WaldEntropy}. But any such change of $a_d^*$ due to higher curvature corrections can be equivalently interpreted as simply a renormalization of the ratio $\lAdS/\lP$ within Einstein gravity. The precise renormalization factor is given by the bracket in \eqref{eq:adHD}. The effect of higher derivative terms is hence not visible for its only imprint is to renormalize a dimensionless parameter of the theory.

Note that the above expressions are divergent and the matchings performed should hence be done more carefully. For example, the volume factor in \eqref{eq:EinsteinEntropy} should be regulated such that its leading divergent piece scales with the area of the sphere $\mathcal{A}$. This divergence depends on the cut-off, but in a consistent way such that the CFT short distance cut-off (regulating the UV divergence of entanglement entropy near the boundary of $\mathcal{A}$) matches the bulk cut-off (regulating the divergent area of $\mathbb{H}_{d-1}$) \cite{Ryu:2006ef}. More importantly, the expression \eqref{eq:EinsteinEntropy} also contains a universal piece which is constant in even dimensions and logarithmically divergent in odd dimensions. For this reason, the above matching conditions strictly speaking refer to the universal piece, which takes the form \cite{Ryu:2006ef,Myers:2010xs}
\begin{equation}
  S_{EE,\text{univ}}^{(0)}(\mathcal{A}) =  \begin{cases} -(-1)^{d/2} \, 4 \, a_d^* \, \log (2R/\delta) & (d\text{ even}) \\
                                              (-1)^{(d-1)/2} \, 2\pi \, a_d^* & (d \text{ odd}) 
                                \end{cases}
\end{equation}
where $\delta$ is the UV cut-off scale in the CFT. It is only for spherical regions $\mathcal{A}$ that our arguments above give the correct result without explicitly dealing with these additional subtleties, see e.g.\ \cite{Hung:2011xb}.

From a CFT point of view, $a_d^*$ is a fundamental parameter. As we have just seen, it characterizes the normalization of the universal part of vacuum entanglement entropy. It is a remarkable fact about entanglement entropy that in even dimensions $a_d^*$ coincides with the coefficient $a$ of the Euler density in the conformal trace anomaly \cite{Schwimmer:2008yh,Myers:2010xs,Myers:2010tj}. In odd dimensions, $a_d^*$ is proportional to $\log Z_{S^d}$, the sphere partition function of the CFT \cite{Casini:2011kv}.

For a detailed reconstruction of bulk dynamics from entanglement, one would like to probe the particular structure of the gravitational action. In this note we wish to study deformations of the above setup and ask what CFT quantities one can use in order to gain sensitivity to features which are peculiar to higher curvature terms in the gravity dual. Clearly the vacuum entanglement entropy of balls $\mathcal{A}$ will not be enough by the above argument. Consider therefore a deformation of the CFT by some relevant scalar operator $\mathcal{O}$, uniformly coupled to the CFT with a coupling $\lambda$. 
Given the arguments above, it is clear what bulk computation we should do in order to learn about the change in $S_{EE}(\mathcal{A})$: add to the gravitational problem a scalar field dual to $\mathcal{O}$ with suitable boundary conditions, solve the scalar equation of motion in the AdS${}_{d+1}$ background and compute the backreaction on the minimal surface area perturbatively in $\lambda$. As we will see, the linearized answer is still universal in a way very similar to how \eqref{eq:WaldEntropy} was universal for the unperturbed problem. Namely, its functional form is completely fixed such that it only depends on a few parameters which encode the influence of higher curvature couplings. These parameters can again be absorbed in a renormalization of dimensionless quantities such that the answer is indistinguishable from what one would have obtained in Einstein gravity minimally coupled to a scalar field. This is the main universality statement studied in this section:
{\it For holographically computing entanglement entropy of spherical regions in CFTs deformed by a relevant scalar operator, any physically acceptable theory of a spin-2 and a spin-0 field with appropriately tuned values of the couplings gives the same result as cosmological Einstein gravity minimally coupled to a scalar. In particular, this quantity does not distinguish between different higher curvature theories of gravity. }
In the rest of this section, we will explain these statements from the CFT and from the gravity side.

\subsection{CFT explanation for universality}
\label{sec:CFTExplanation}

We start by reviewing from a CFT point of view the ingredients involved in computing entanglement entropy of spherical regions in perturbed states. Our discussion closely follows \cite{Faulkner:2014jva,Rosenhaus:2014woa,Rosenhaus:2014zza}.

Consider a CFT on $d$-dimensional Euclidean flat space and perform the following deformation of the CFT action $\In_{CFT}$ by a relevant scalar operator $\mathcal{O}$ of dimension $\Delta<d$: 
\begin{equation}
 I_{CFT} = \In_{CFT} + \lambda \int d^dx \; \mathcal{O}(x)  \,.
\end{equation}
We can compute the perturbative changes of the entanglement entropy $S_{EE}(\mathcal{A})$ by writing the reduced density matrix as a path integral. This path integral lives on a manifold which is the Euclidean spacetime with a cut along $\partial \mathcal{A}$. The matrix element corresponding to some field configurations $\phi_+$ and $\phi_-$ on the two sides of the cut is
\begin{equation}
\begin{split}
 \langle \phi_- | \rho_\mathcal{A}| \phi_+ \rangle &= \int_{\small\substack{\phi(\partial\mathcal{A}_+) = \phi_+ \\ \phi(\partial\mathcal{A}_-) = \phi_-}} \,[\mathcal{D}\phi] \; e^{-I_{CFT}}\\
 &= \int_{\small\substack{\phi(\partial\mathcal{A}_+) = \phi_+ \\ \phi(\partial\mathcal{A}_-) = \phi_-}} \,[\mathcal{D}\phi] \; e^{-\In_{CFT}} \left( 1 - \lambda \int d^dx\, \mathcal{O}(x) + \frac{\lambda^2}{2} \iint d^dx\,d^dy\, \mathcal{O}(x) \mathcal{O}(y) + \ldots \right) \,. 
\end{split}
\end{equation}
Now define a modular Hamiltonian $H_\mathcal{A}$ via $\rho_\mathcal{A} = e^{-H_\mathcal{A}}$. We can then use the identity $S_{EE}(\mathcal{A}) = \text{Tr}(\rho_\mathcal{A} \, H_\mathcal{A})$ to compute the change in the entanglement to any desired order in $\lambda$:\footnote{ Operators without argument are understood to be integrated, e.g., $\mathcal{O} \equiv \int d^dx \; \mathcal{O}(x)$.}
\begin{equation}\label{eq:PertSEE}
\begin{split}
 S_{EE}(\mathcal{A}) 
 &= \Sn_{EE}(\mathcal{A}) + \lambda \, \left[\frac{\partial S_{EE}(\mathcal{A})}{\partial \lambda}\right]_{\lambda = 0} + \frac{\lambda^2}{2} \, \left[\frac{\partial^2 S_{EE}(\mathcal{A})}{\partial \lambda^2} \right]_{\lambda=0} + \ldots \\
 &= \Sn_{EE}(\mathcal{A}) + \lambda \left[ -\langle \mathcal{O} H_\mathcal{A} \rangle + \langle \frac{\partial H_\mathcal{A}}{\partial \lambda} \rangle \right]_{\lambda=0} + \frac{\lambda^2}{2} \left[\langle \mathcal{O} \mathcal{O} H_\mathcal{A} \rangle - \langle \mathcal{O} \frac{\partial H_\mathcal{A}}{\partial \lambda} \rangle + \langle \frac{\partial^2 H_\mathcal{A}}{\partial^2 \lambda} \rangle\right]_{\lambda=0} + \ldots \\
 &= \Sn_{EE}(\mathcal{A}) - \lambda \langle \mathcal{O} \Hn \rangle  + \frac{\lambda^2}{2} \left( \langle \mathcal{O} \mathcal{O} \Hn \rangle - \langle \mathcal{O} \frac{\partial H_\mathcal{A}}{\partial \lambda} \rangle\right) + \ldots 
 \end{split}
\end{equation}
where $\Sn_{EE}(\mathcal{A})$ is universal up to an overall normalization $a_d^*$ as explained in \S\ref{sec:setup}, and similarly $\Hn$ is the unperturbed modular Hamiltonian. In the last step we used the fact that $\rho_\mathcal{A}$ is normalized which implies $\langle \frac{\partial H_\mathcal{A}}{\partial \lambda} \rangle = 0$.
We can now use that $\mathcal{A}$ is a spherical region, which brings about another important simplification: the modular Hamiltonian can be traded for an integrated stress tensor, since for spherical $\mathcal{A}$ with radius $R$ centered at the origin, one finds
\begin{equation}\label{eq:ModHamDef}
 H_\mathcal{A} = 2\pi \int_{\mathcal{A}} d^{d-1}x \; \frac{R^2-\vec{x}\,{}^2}{2R} \, T_{00}(x) \,,
\end{equation} 
for $\mathcal{A}$ lying on the $x^0 = 0$ slice. In our problem, correlation functions involving the modular Hamiltonian can therefore be regarded as correlation functions involving the conformal stress tensor. But for stress tensor correlation functions, we can use well-known features of CFTs:
on general grounds, two-point correlations between a stress tensor and a primary operator are constrained by conformal invariance and conformal Ward identities \cite{Erdmenger:1996yc}. In fact, independent of the details of our problem, conformal invariance of a theory in a flat spacetime enforces $\langle \mathcal{O}T_{\mu\nu} \rangle = 0$. Therefore the $\order(\lambda)$ perturbation in \eqref{eq:PertSEE} vanishes.\footnote{ We will use indices $a,b,\ldots$ for bulk coordinates, $\mu,\nu,\ldots$ for boundary directions, and $\alpha,\beta,\ldots$ for intrinsic coordinates on the minimal surface $\mathcal{E}_{min}$.} 

Furthermore, the perturbation of $T_{\mu\nu}$ is uniform in $\mathcal{O}$ and we have $\frac{\partial T_{\mu\nu}}{\partial\lambda} = -  \delta_{\mu\nu}\,\mathcal{O}$. To see this, recall that for a Euclidean flat boundary, we have
\begin{equation}\label{eq:CFTstress}
  T_{\mu\nu}(x) = \frac{2}{\sqrt{g}} \frac{\delta I_{CFT}}{\delta g^{\mu\nu}(x)} = T_{\mu\nu}^{\text{\tiny (0)}}(x) - \delta_{\mu\nu} \, \lambda\, \mathcal{O}(x) \,.
\end{equation}
The entanglement entropy in the deformed theory, \eqref{eq:PertSEE}, hence takes a generic form \cite{Rosenhaus:2014zza}:
\begin{equation}\label{eq:PertSEE2}
\begin{split}
 S_{EE}(\mathcal{A}) 
  &= \Sn_{EE}(\mathcal{A}) + \frac{\lambda^2}{2} \left( \langle \mathcal{O} \mathcal{O} \Hn \rangle - \langle \mathcal{O} \mathcal{O} \rangle\right) + \ldots 
 \end{split}
\end{equation}
As we can see, this is only sensitive to the correlators $\langle \mathcal{O} \mathcal{O} T_{\mu\nu} \rangle$ and $\langle \mathcal{O} \mathcal{O} \rangle$. These, however, are known to be universal functions in any conformal field theory which only depend on one overall normalization constant:
\begin{equation}\label{eq:OOnormalization}
\begin{split}
 \langle \mathcal{O}(x_1)\mathcal{O}(x_2) \rangle = \frac{ \cOO}{(x_1-x_2)^{2\Delta}} 
 \end{split}
\end{equation}
and a similar universal function for $\langle \mathcal{O}(x_1) \mathcal{O}(x_2) T_{\mu\nu}(x_3) \rangle$ whose normalization is also determined solely by $\cOO$ \cite{Osborn:1993cr,Erdmenger:1996yc}. The entanglement entropy of $\mathcal{A}$ is hence completely universal up to $\order(\lambda^2)$, only depending on the fundamental constants $\{a_d^*\,,\,\cOO\}$ and the operator dimension $\Delta$. 

The divergent structure of the expression \eqref{eq:PertSEE2} has been computed in \cite{Rosenhaus:2014zza} just using standard Euclidean CFT techniques. However, we must now face an additional subtlety: the analysis of \cite{Faulkner:2014jva}, which treats the problem at hand using the replica trick and carefully analyzes the analytic structure of the resulting partition functions, leads to a different result which involves a non-trivial finite piece. Nevertheless the basic lesson of this more involved analysis using the replica trick turns out to be still the same; namely the $\order(\lambda^2)$ perturbation of entanglement entropy is determined solely by correlators  $\langle\mathcal{O}\mathcal{O}\rangle$ and $\langle\mathcal{O}\mathcal{O}T_{\mu\nu}\rangle$. To be precise, the finite contribution to the correction at $\order(\lambda^2)$ reads \cite{Liu:2012eea,Nishioka:2014kpa,Faulkner:2014jva}:\footnote{ This is assuming that $\mathcal{O}$ is a relevant operator with $\Delta \neq d/2$. See \cite{Liu:2012eea,Nishioka:2014kpa} for a discussion of the case $\Delta = d/2$ in holographic theories.}
\begin{equation} \label{eq:SEECFT}
 S_{EE}(\mathcal{A}) = \Sn_{EE}(\mathcal{A}) - \lambda^2 R^{2(d-\Delta)}\, \frac{\pi^{\frac{d+1}{2}}(d-\Delta) \Gamma(1+\tfrac{d}{2} - \Delta)}{2\,\Gamma(\tfrac{3}{2} + d - \Delta)} + \ldots  \,,
\end{equation}
where $R$ is the radius of the ball $\mathcal{A}$.
In this expression the normalization $\cOO$ has been fixed in a way that will prove convenient when comparing with holographic calculations. However, we will not need the explicit form of $S_{EE}(\mathcal{A})$ for the following analysis.

\subsection{Gravitational explanation for universality}
\label{sec:GravExplanation}

Having seen that the second order computation of perturbed entanglement entropy of spherical regions is only sensitive to $\langle \mathcal{O}\mathcal{O}\rangle$ and $\langle  \mathcal{O}\mathcal{O}T_{\mu\nu}\rangle$ correlators (which are universal up to one normalization constant), we now want to explain what this means in gravity.

\subsubsection{Holographic parameter matching}
\label{sec:HoloMatching}

A major hint for understanding the entanglement universality from a holographic point of view is the result of \cite{Faulkner:2014jva}, where it is shown that for the CFT deformation characterized by a coupling $\lambda$ and an operator dimension $\Delta$ as described in \S\ref{sec:CFTExplanation}, one can calculate the entanglement entropy at $\order(\lambda^2)$ by solving an auxiliary gravitational problem described by the action  
\begin{equation} \label{eq:GravAux}
\begin{split}
  I_{aux} = \frac{1}{2\lP^{d-1}}\int \sqrt{-g} \, \left[R + \frac{d(d-1)}{\ell^2} 
      - \frac{\lphi}{2}\left( (\nabla \phi)^2 + m^2 \, \phi^2 \right)\right] \,.
      \end{split}
\end{equation}
The AdS${}_{d+1}$ background geometry $\gn_{ab}$ has a scale which coincides with the cosmological scale: $\lAdS^2 = \ell^2$. Further, $\lphi$ parametrizes the normalization of the scalar action. The prescription is to solve first the scalar wave equation in the background $\gn_{ab}$ and then compute its backreaction on the metric at next order in perturbation theory to get the change of the area of the Ryu-Takayanagi minimal surface. For this task we are therefore not interested in the full action \eqref{eq:GravAux}, but only in the linearized equations of motion due to a scalar probe in AdS${}_{d+1}$ background. In the following all objects with a bar (e.g., $\bar{\nabla}_a$) refer to the background metric $\gn_{ab}$.

We will now comment on the number of free parameters in this problem and how to determine them, given that they should induce the CFT parameters at the boundary in a way consistent with AdS/CFT. A priori there are three dimensionless quantities $\{(\lAdS/\lP)\,,\,\lAdS^2m^2,\lphi\}$ in the action \eqref{eq:GravAux}.
The scalar wave equation in AdS, $(\Boxn-m^2)\phi= 0$, has two asymptotic solutions with respective falloff behavior $z^{\Delta_+}$ and $z^{\Delta_-}$, where the boundary is at $z=0$ (in Poincar\'e coordinates such as \eqref{eq:coords}) and 
\begin{equation}
\Delta_\pm \equiv \frac{d}{2} \pm \sqrt{\frac{d^2}{4} + \lAdS^2 \, m^2} \,.
\end{equation} 
The boundary condition for the scalar is required to be such that it sources an operator with given dimension $\Delta$ and coupling $\lambda$. By standard AdS/CFT techniques \cite{Aharony:1999ti}, we therefore assume\footnote{ The assumption $\Delta \geq d/2$ corresponds to performing so-called standard quantization. Note that $(d-2)/2 \leq \Delta \leq d/2$ would still be compatible with field theory unitarity. The latter regime can be treated by choosing $\Delta_- = \Delta$. See \cite{Hung:2011ta} for a discussion of this case.} $\Delta \geq d/2$, tune the scalar mass such that $\Delta_+ = \Delta$ and fix the dominant asymptotics as $\phi(z\rightarrow 0) \sim \lambda \, z^{\Delta_-}$. 
The mass of the scalar is also immediately fixed by demanding that it sources an operator with dimension $\Delta$; the usual prescription for this is to take $-d^2/4 < \lAdS^2\,m^2<0$ such that
\begin{equation}\label{eq:MassRelation}
\lAdS^2 \, m^2 = \Delta (\Delta-d) \,.
\end{equation}

The remaining two parameters of the theory also need to be fixed by matching with the CFT. On the one hand, $\lAdS/\lP$ makes an appearance in the normalization of the universal scaling of entanglement in the CFT in terms of the generalization of $a$-central charge, c.f., \eqref{eq:WaldEntropy} and can hence be fixed by demanding \eqref{eq:adEH}. On the other hand, $\lphi$ parametrizes rescalings of the bulk field $\phi$ in the same sense in which the dual boundary two-point function $\langle\mathcal{O}\mathcal{O}\rangle$ depends on an overall normalization $\cOO$ as in \eqref{eq:OOnormalization}. 
Since $\phi$ has already been equipped with boundary conditions which lead to the correct coupling $\lambda$ for $\mathcal{O}$, we can follow standard techniques to work out the normalization $\lphi$ in terms of $\cOO$. This involves computing the bulk to boundary propagator for $\phi$ and matching its boundary two-point function with the CFT correlator $\langle\mathcal{O}\mathcal{O}\rangle$. The result is \cite{Aharony:1999ti,Hung:2011ta}
\begin{equation}\label{eq:cOOmatch}
 \cOO = \frac{\Delta-d/2}{\pi^{d/2}} \frac{\Gamma(\Delta)}{\Gamma(\Delta-d/2)} \frac{\lAdS^{d-1}}{\lP^{d-1}}\;\lphi \,.
\end{equation}

We stress that, as shown in \cite{Faulkner:2014jva}, the auxiliary gravitational problem just described can be used for computing perturbed entanglement entropy of ball shaped regions in {\it any} CFT.
Given the universality of the correlators $\langle \mathcal{O}\mathcal{O}\rangle$ and $\langle \mathcal{O}\mathcal{O}T_{\mu\nu}\rangle$ in all CFTs (including those which are dual to cosmological Einstein gravity), this result is not very surprising. It is, however, quite remarkable that the most natural way of rewriting the CFT problem is in terms of the auxiliary gravitational system \eqref{eq:GravAux}. As we shall see, there are many other (higher curvature) models of gravity which could equally well serve as the auxiliary system for this problem. Turning this logic around, all these different gravitational theories are indistinguishable if only perturbed entanglement data of spheres is measured. We now turn to illuminating this point from the gravity side.

\subsubsection{Construction of the most general gravitational model}
\label{sec:GravConstruction}

We now want to make more precise the statement that all physically acceptable gravity models are equivalent for the computation at hand. Our approach is to construct a large class of auxiliary gravitational systems which make the same predictions as \eqref{eq:GravAux} in this respect. 
Let us take the CFT data as given and try to construct a holographic description in terms of semiclassical gravity 'bottom up'. The first step of such a program could be the question: what are the minimal bulk ingredients if they are supposed to compute entanglement entropy of spheres in vacuum and after turning on the perturbation? 
We will now answer this question by building general linearized bulk actions which satisfy the following conditions:
\begin{itemize}
\item The field content is a symmetric spin-2 tensor (the metric) and a scalar field, which source the boundary operators $T_{\mu\nu}$ and $\mathcal{O}$, respectively. The model gives dynamics to these fields and defines a coupling between them which allows to compute the linearized backreaction $h_{ab}$ on an AdS${}_{d+1}$ background $\gn_{ab}$ after perturbing the latter with the scalar probe $\phi$. 
\item There are no ghost degrees of freedom; in particular the linearized equations of motion on AdS${}_{d+1}$ are second order in derivatives. (See below for a justification of this assumption and \S\ref{app:generalization} for generalizations in the case where the higher curvature terms are perturbatively small and this assumption can be dropped.)
\item The action is diffeomorphism invariant (c.f., \cite{Ogievetsky:1965, PhysRevD.33.3613}). 
\end{itemize}
We will find that these conditions specify the linearized action uniquely up to normalization constants; see \eqref{eq:GeneralAction2ndOrder} for the final result.

Let us briefly explain why the equations of motion should be second order. 
It is well known that equations of motion which are higher than second order lead to pathological behaviour which we want to exclude from our discussion of theories which are dual to unitary CFTs. In particular, higher order equations of motion generically give rise to further propagating degrees of freedom with negative energy \cite{Stelle:1977ry,Stelle:1977aa,Barth:1983aa,Biswas:2011ar,Conroy:2015wfa}.\footnote{ \label{footnote:fR} There are special circumstances where the additional degrees of freedom can be dealt with by using appropriate field redefinitions. For instance, $f(R)$ theories of gravity naively have higher order equations of motion, but are known to be equivalent under field redefinitions to Einstein gravity coupled to a (positive energy) scalar field such that the boundary theory is unitary \cite{Sotiriou:2008rp}. The graviton in this case couples not only to the boundary stress tensor, but in addition to some scalar operator. The CFT therefore has some more specific features than we assumed in the beginning. For validating our conclusions  in the most general circumstances, it would be interesting to explore the most general requirements for a bulk theory to admit a unitary dual CFT. A first step in this direction might be a gauge-fixed analysis where the scalar part of the graviton is set to zero, c.f., the transverse traceless gauge employed in \S\ref{app:generalization}. This would at least take care of $f(R)$ theories.} The presence of ghosts in such theories thus leads to classical instabilities \cite{Boulware:1985nm} (which would lead to negative norm states and therefore non-unitary quantum theories upon quantization). These features would also be visible in the dual CFT as operators with complex conformal dimension or negative norm. 
Indeed, via holography the problem of perturbed entanglement entropy is described in the CFT by correlation functions of the stress tensor $T_{\mu\nu}$ and a relevant scalar operator $\mathcal{O}$. This setup should be described in gravity by nothing more than a spin-2 graviton which sources $T_{\mu\nu}$ and a scalar field to source $\mathcal{O}$.\footnote{ Of course, it would be possible for $\mathcal{O}$ to be some other scalar operator, e.g., one corresponding to the square of a scalar field. Since such scenarios would not change our analysis qualitatively, we will assume that $\mathcal{O}$ is simply sourced by $\phi$ with appropriate boundary condition.} If the graviton equations of motion were higher than second order in derivatives, then the additional unwanted ghost modes would couple to other (non-unitary) operators. Therefore, whatever the full gravitational theory is, after linearization of the action around AdS${}_{d+1}$ the excitation $h_{ab}$ is bound to appear in such a way as to yield second order equations of motion. 

Despite these arguments in favour of two-derivative equations of motion, higher derivative models with higher order equations of motion are regularly studied as toy models for semiclassical corrections to Einstein gravity induced by some consistent truncation of full unitary quantum gravity (which is assumed to be UV complete). Such models can be justified as sensible toy models in a perturbative framework as long as the ghost modes are very heavy and do not go on-shell (which turns out to be the case whenever the higher derivative couplings are small). We explore this setup in \S\ref{app:generalization}, where we show that in this context our arguments concerning entanglement entropy still hold. The reason for the latter is that at low energies the ghosts can be ignored and the sector of physical modes is exactly the same as for ghost-free models which we are now going to analyze.

Let us now construct the minimal ingredients for a linearized action for $h_{ab}$ and $\phi$,  which leads to second order equations of motion. We will argue that the most general action of this type is the same as the linearization of Einstein-scalar theory \eqref{eq:GravAux} up to various normalizations. For illustration, we can hence start by expanding the metric around an AdS${}_{d+1}$ background solution $\gn_{ab}$ as $g_{ab} = \gn_{ab} + h_{ab}$ and linearizing the Einstein action \eqref{eq:GravAux}. Including some convenient normalization factors, this procedure leads to\footnote{ To verify this by hand, the identities in \S\ref{sec:perturbation} are helpful.}
\begin{equation} \label{eq:GeneralAction2ndOrder}
\begin{split}
  I = \frac{1}{2\lP^{d-1}}\int\sqrt{-\gn} \; \Bigg\{   
    & -\frac{2d}{\lAdS^2}\frac{\Gamma(d/2)}{\pi^{d/2}}\frac{\lP^{d-1}}{\lAdS^{d-1}} \, a_d^* + \frac{\lh}{2}\, h_{ab}\, \prop^{abcd}\,h_{cd} +   \lphi\,\left[ -\frac{1}{2} \left( (\dn\phi)^2 + m^2 \phi^2 \right)\right]  \\
     & 
     + \frac{\lphi}{2}\, h^{ab} \left( - \frac{1}{2} \left[(\dn\phi)^2 + m^2 \phi^2\right]\gn_{ab} + \dn_a \phi \,\dn_b \phi \right)+ \order(h^3,h^2\phi^2)\Bigg\} \,,
\end{split}
\end{equation}
where $\prop_{abcd}$ is the graviton kinetic operator
\begin{equation} \label{eq:EHlichnerowicz}
\begin{split}
 \prop_{abcd}&=
     \frac{1}{2}\left(\gnt_{a(c} \gnt_{d)b}- \gnt_{ab} \gnt_{cd}\right) \, \Boxn - \dn_{(c} \gnt_{d)(a}\dn_{b)}+ \frac{1}{2} \left( \gnt_{ab} \dn_c \dn_d + \gnt_{cd} \dn_a \dn_b \right) \\
                  &\quad + \frac{d}{\lAdS^2}\left(\frac{1}{2}\,\gnt_{ab} \gnt_{cd} - \gnt_{a(c} \gnt_{d)b}\right) \,.
\end{split}
\end{equation}
A few explanations are in order concerning the way we wrote \eqref{eq:GeneralAction2ndOrder}:\\
(\it{i$\,$}\normalfont) In writing \eqref{eq:GeneralAction2ndOrder} we introduced a normalization $\lh$ in the graviton kinetic term for later convenience. By expanding the Einstein-scalar action \eqref{eq:GravAux}, one finds simply $(\lh)_{Einstein} = 1$, but we will shortly encounter other (higher derivative) theories with other values of $\lh$.\\
(\it{ii$\,$}\normalfont) Similarly, we wrote a constant term proportional to $a_d^*$. This is also using the benefit of hindsight (in particular \eqref{eq:adHD}) to already implement the correct generalization for higher derivative theories. For the present case of Einstein gravity, $(a_d^*)_{Einstein}$ is just given by \eqref{eq:adEH}. For more general theories, notice that the constant term in \eqref{eq:GeneralAction2ndOrder} is determined by the requirement that switching off $h_{ab}$ and $\phi$ in the action is equivalent to evaluating the full non-linear Lagrangian on the AdS background. By means of \eqref{eq:adHD} we know that the constant thus obtained needs to be $a_d^*$ (with appropriate pre-factors).\\ 
(\it{iii$\,$}\normalfont) Note also that we did not write the $\order(h)$ term of the action. This term would just define the equations of motion for the AdS${}_{d+1}$ background geometry $\gn_{ab}$. By assuming an expansion of the form $g_{ab} = \gn_{ab} + h_{ab}$, such terms vanish automatically.\\

The graviton-scalar action \eqref{eq:GeneralAction2ndOrder} was obtained by linearizing the Einstein-scalar system \eqref{eq:GravAux} around a given AdS background with scale $\lAdS$. However, we will now argue that {\it any} theory which satisfies the basic conditions listed at the beginning of this subsection, takes this form. As we will see, the only freedom is parameterized by the normalizations $\{a_d^*, \lh, \lphi, m^2\}$. Of these, $a_d^*$ has already been fixed manifestly by using \eqref{eq:adHD}. Also $\lphi$ and $\lAdS^2\,m^2$ could be fixed in terms of field theory quantities via \eqref{eq:MassRelation} and \eqref{eq:cOOmatch}. The graviton normalization, $\lh$, will be discussed shortly.  
This will then demonstrate the universality of the dynamics of perturbing an AdS geometry with a scalar field up to normalizations.

\paragraph{Graviton propagator on AdS${}_{d+1}$:}
Clearly, the field content in the bulk should involve a spin-0 field $\phi$ to probe the geometry, whose boundary condition $\phi(z\rightarrow 0) \sim \lambda \, z^{d-\Delta}$ sources the relevant deformation $\mathcal{O}$. But before specifying the matter probe in this much  detail, let us discuss the general gravitational response to any small perturbation. This is described by a spin-2 field propagating on the background $\gn_{ab}$. We will write this as $g_{ab} = \gn_{ab} + h_{ab}$, where $h_{ab}$ is the linearized perturbation. 

We now want to establish the most general physically acceptable equations of motion for $h_{ab}$, as described above. 
It is straightforward to write the most general (up to total derivatives) action at $\order(h^2)$ of the symmetric spin-2 field which involves at most two derivatives. We can parametrize it as (see, e.g., \cite{Alvarez:2006uu})
\begin{equation} \label{eq:2ndOrderLin}
\begin{split}
  I_{(h^2)}  =\frac{1}{2\lP^{d-1}}\int \sqrt{-\gn} \;  \bigg\{ -\frac{\lh}{4} \dn_a h_{bc} \dn^a h^{bc} & + \frac{b_1}{2} \dn_b h^{ac} \dn_a h^b_c - \frac{b_2}{2} \dn^a h \dn^b h_{ab} \\
  & + \frac{b_3}{4} \dn_a h \dn^a h  
  +  \frac{M^2}{2} \left( h^2 - b_4 \,h_{ab} h^{ab} \right) \bigg\} \,, 
\end{split}
\end{equation}
where $h \equiv h^{ab} \gn_{ab}$. All raising and lowering of indices as well as covariant derivatives with a bar refer to $\gn_{ab}$.\footnote{ Note that the coefficient $\lh$ of the first term in \eqref{eq:2ndOrderLin} cannot be vanishing as this would mean that no spin-2 degrees of freedom propagate at all. By this we ensure that no gauge choice can eliminate the spin-2 mode. For instance, we could choose to work in a transverse traceless gauge ($\dn_a h^{ab} = 0 = h$), but the first term would still survive as it is the source of the spin-2 component. 
On a similar note, we need $\lh > 0$ in order for the spin-2 graviton not to be ghost.} 
We can already see that compared to the infinite number of possible higher derivative theories of gravity, the above linearized action is quite restrictive, having only six free parameters $\{\lh,M^2,b_1,b_2,b_3,b_4\}$. In fact the action will be even more constrained once we impose that the full theory should exhibit diffeomorphism invariance. It is a classic result that demanding a diffeomorphism invariant theory beyond the linearized level singles out a unique graviton propagator \cite{VanNieuwenhuizen:1973fi}. In our explicit notation, we can give a simple justification of this statement by considering a spin-2 gauge transformation (aka linearized diffeomorphism) $h_{ab} \rightarrow h_{ab} + 2\partial_{(a} \xi_{b)}$. 
The $\order(h^2)$ action transforms nontrivially:
\begin{equation}
\begin{split}
\delta I_{(h^2)} = \frac{1}{2\lP^{d-1}}\int  \sqrt{-\gn} \; \xi^a  \bigg\{& (b_1-\lh)  \dn^b \Boxn h_{ab} + (b_1-b_2)   \dn_a \dn_c \dn_d h^{cd}  + (b_3-b_2)   \dn_a \Boxn  h \\
& - \left(2M^2 - \frac{d\,b_2}{\lAdS^2} \right) \dn_a h + \left(2b_4\, M^2 - \frac{2\,d\,b_1}{\lAdS^2}\right) \dn_b h^b_a \bigg\}\,,
\end{split}
\end{equation}
where we used partial integration and \eqref{eq:AdSsimplify}.
In order for the action to be diffeomorphism invariant, the above expression has to vanish for arbitrary $\xi^a$. Since all terms are independent, this condition forces upon us five relations among the six parameters:
\begin{equation}
b_1 = b_2 = b_3 = \lh  \,,\qquad M^2 = \frac{d}{2\lAdS^2}\,\lh \,,\qquad b_4 = 2\,.
\end{equation}
Any other choice of parameters would not correspond to the linearization of a diffeomorphism invariant theory.
The only free parameter which is left undetermined is the overall scale $\lh$ of the graviton action. In a full non-linear theory of gravity this is, of course, fixed by the normalization of the action and we will compute some examples in \S\ref{sec:exampleGB} to illustrate this.
For instance, we have already seen that the  graviton action \eqref{eq:2ndOrderLin} is exactly the same as the second order term in a perturbative expansion of the Einstein-Hilbert action with cosmological constant scale $\ell = \lAdS$ and overall normalization $\lh = 1$.

This construction shows that by just looking at the linearized graviton propagator on AdS of any reasonable diffeomorphism invariant theory of gravity, we will never be able to distinguish the theory from cosmological Einstein gravity. The underlying reason for this is the universality of spin-2 propagation on maximally symmetric backgrounds.

\paragraph{Coupling to scalar matter:}
Having constructed the most general quadratic action for the metric perturbation, we should now in a similar fashion argue that \eqref{eq:GeneralAction2ndOrder} couples $h_{ab}$ to the scalar field in the most general way at this order. There are several possible Yukawa couplings and self-interaction terms. However, for consistency of our perturbative scheme, the only terms we are interested in at this order are $\phi\phi$ self-interactions and $h\phi\phi$ couplings. We will now describe the most general actions of these kinds:
\begin{itemize}
\item {\bf Scalar field kinetic and mass term:}
At lowest order in perturbation theory (the probe limit), we are only concerned with couplings of $\phi \phi$ to the background geometry. 
Up to total derivatives, the most general such action with at most two time derivatives may be parametrized in terms of a mass parameter $m^2$ and an overall dimensionless normalization $\lphi$:
\begin{equation}\label{eq:ScalarAction2}
I_{(\phi^2)} =\frac{1}{2\lP^{d-1}} \int \sqrt{-\gn}\; \lphi \, \left\{ -\frac{1}{2} \left( (\dn\phi)^2 + m^2 \phi^2 \right)\right\}\,.
\end{equation} 
This action defines the background dynamics of $\phi$ as $(\Boxn - m^2) \phi = 0$. 

\item {\bf Backreaction on the geometry:} 
There are several possible couplings of the type $h\phi\phi$, which are inequivalent upon integration by parts. We can parametrize them as follows:
\begin{equation} \label{eq:hphiphi}
\begin{split}
 I_{(h\phi^2)} = \frac{1}{2\lP^{d-1}}\int \sqrt{-\gn} \;  \bigg\{ -\frac{h}{4} \big( c_1 (\dn\phi)^2 &+ c_2 \, m^2 \phi^2 \big) + \frac{c_3}{2}\, h^{ab} \dn_a \phi \dn_b \phi \\
 & + c_4\, h^{ab} \phi \dn_a \dn_b \phi + c_5\, h \phi \Boxn \phi \bigg\} \,,
\end{split}
\end{equation}
%
Clearly the coefficients $c_2$ and $c_5$ are on-shell equivalent. We will thus set $c_5 = 0$ in what follows. 
Now demanding diffeomorphism invariance, an argument completely analogous to the one we gave for \eqref{eq:2ndOrderLin} implies the following conditions for the remaining parameters:\footnote{ Alternatively, one can determine the coefficients by demanding conservation of the matter stress tensor, $\dn^b T_{ab}^{(\phi)} = 0$.}
\begin{equation}\label{eq:constraints1}
 c_1 = c_2 = c_3   \,, \quad c_4 = 0 \,. 
\end{equation}
which leaves us with only one free parameter $c_1$. However, it is immediately clear that the overall normalization has to be the same as in \eqref{eq:ScalarAction2}, i.e., 
\begin{equation}
c_1 = \lphi \,.
\end{equation}
If this normalization was any different, $h_{ab}$ would not be the response to a scalar with mass $m^2$ and normalization $\lphi$. 
Taking into account these constraints, we can rewrite \eqref{eq:hphiphi} as
\begin{equation}\label{eq:Tphi}
\begin{split}
 I_{(h\phi^2)} &= \frac{1}{2\lP^{d-1}}\int \sqrt{-\gn} \; \frac{\lphi}{2} \, h^{ab}\, T_{ab}^{(\phi)} \,,\\
  \text{with}\quad T_{ab}^{(\phi)} &\equiv - \frac{1}{2} \left[(\dn\phi)^2 + m^2 \phi^2\right]\gn_{ab} + \dn_a \phi \,\dn_b \phi \,,
\end{split}
\end{equation}

\end{itemize}

Summarizing these two points, we recognize the most general scalar-graviton couplings which are consistent with the perturbative scheme and diffeomorphism invariance as being exactly the same as the expansion of the following action up to $\order(h)$:
\begin{equation}\label{eq:Matteraction}
\begin{split}
 I_{matter} &= I_{(\phi^2)} + I_{(h\phi^2)} + \order(h^2) \\
 &=\frac{1}{2\lP^{d-1}} \int \sqrt{-g} \; \lphi \, \left[ -\frac{1}{2} \left( (\nabla\phi)^2 + m^2 \phi^2 \right)\right] +{\order(h^2)} \,,
\end{split}
\end{equation}
where $g_{ab} = \gn_{ab} + h_{ab}$. 
Of course, this result is not too surprising as this is the standard action for a massive scalar minimally coupled to gravity. 

\paragraph{Most general second order action: }
We can summarize the above discussion as follows. The most general way of coupling $h_{ab}$ to a scalar $\phi$ in a way that is consistent with our perturbative scheme and respects diffeomorphism invariance and some very basic constraints coming from the particular CFT computation we are interested in, is given by \eqref{eq:GeneralAction2ndOrder}.
This action has only four dimensionless parameters $\{\lAdS/\lP \,,\, m^2\lAdS^2\,, \lphi\,,\lh\}$. 
We already explained in \S\ref{sec:HoloMatching} how for our problem the first three of these have to be determined holographically in terms of the dimensionless CFT quantities $\{a_d^*\,,\,\Delta\,,\,\cOO\}$. We will argue shortly that the fourth parameter $\lh$ does not enter the computation of interest and therefore its value is irrelevant for computing perturbed entanglement entropy. In addition, there is, of course, the boundary coupling $\lambda$ which enters the gravitational setup by prescribing the boundary condition for $\phi$. 

As a consequence, any covariant theory of gravity whose linearization takes the form \eqref{eq:GeneralAction2ndOrder}, provides a good auxiliary system for computing $S_{EE}(\mathcal{A})$ at second order in perturbation theory. 
For instance, the action \eqref{eq:GeneralAction2ndOrder} with $\lh = 1$ is precisely the same as the linearization of the cosmological Einstein-scalar action \eqref{eq:GravAux} via $g_{ab} = \gn_{ab} +  h_{ab}$.
There are, of course, many other non-linear actions whose linearization on maximally symmetric backgrounds also takes the form \eqref{eq:GeneralAction2ndOrder} with a different value of $\lh$, e.g., those of Lovelock type \cite{Lovelock:1971yv} or quasi-topological gravity \cite{Myers:2010ru}. Indeed, even if the CFT in question is actually dual to some particular higher derivative theory; for the purpose of calculating the perturbed entanglement entropy, one can always work with the Einstein action \eqref{eq:GravAux} after suitably matching the parameters $\{\lAdS/\lP \,,\, m^2\lAdS^2 \,, \lphi\}$ with the CFT boundary conditions. The different linearized actions will then only differ by their value of $\lh$. We will see examples of this in \S\ref{sec:exampleGB}. 

Since the normalization $\lh$ is the only ingredient in \eqref{eq:GeneralAction2ndOrder} which would genuinely distinguish bulk theories with higher derivative couplings from those without, we now need to understand why the linearized computation of perturbed entanglement entropy of a ball is not sensitive to this. 

\paragraph{Holographic interpretation of $\lh$:}
Before we continue with the gravitational analysis, it is instructive to elaborate on the meaning of $\lh$, which has a very natural interpretation from the CFT point of view.
The graviton normalization $\lh$ corresponds to the normalization of the CFT stress tensor two-point function, usually denoted by $\CT$. 
It sets the strength of the leading divergence in the two-point function of CFT stress tensors:
\begin{equation}\label{eq:TTcorr}
 \langle T_{\mu\nu}(x) T_{\rho\sigma}(0) \rangle = \frac{\CT}{x^{2d}} \, \mathcal{I}_{\mu\nu,\rho\sigma}(x) \,,
\end{equation}
with some well known universal function $\mathcal{I}_{\mu\nu,\rho\sigma}(x)$ \cite{Osborn:1993cr,Erdmenger:1996yc}. In $d=2$, the normalization agrees with the usual $c$-central charge, $\CT=c$. Similarly, in $d=4$ one finds $\CT = (40/\pi^2) c$. Once a holographic description of the CFT has been given, usual AdS/CFT methods can be used to compute the left hand side of \eqref{eq:TTcorr} as a bulk graviton correlator. Since $\lh$ sets the normalization of the graviton propagator in the presence of matter sources (c.f., \eqref{eq:generalEOM0}), it is rather clear that $\CT$ will be proportional to $\lh$. Doing the holographic calculation in detail leads to the standard result \cite{Buchel:2009sk}
\begin{equation}\label{eq:cTmatch}
 \CT = \frac{f_d}{2}\, \frac{\lAdS^{d-1}}{\lP^{d-1}} \,\lh \,,\quad \text{where}\quad f_d \equiv \frac{2d(d+1)}{d-1} \frac{\Gamma(d)}{\pi^{d/2}\Gamma(d/2)} \,.
\end{equation}
It is now obvious that once $\lAdS/\lP$ has been fixed such as to reproduce the other central charge $a_d^*$ via matching \eqref{eq:adHD}, there is no free parameter left in \eqref{eq:cTmatch} which could absorb the normalization $\lh$. In our setup, gravity theories with different $\lh$ will hence describe CFTs with different $\CT$. Conversely, a CFT measurement of $\CT$ would allow to for a distinction between different bulk models. However, we will now argue that the perturbed entanglement entropy of spheres is agnostic about this.

\subsubsection{Second order Wald entropy is independent of graviton normalization }
\label{sec:lhindependence}

As we have seen, any gravity theory whose action linearized around AdS${}_{d+1}$ takes the form \eqref{eq:GeneralAction2ndOrder} can be used to compute second order perturbations of entanglement entropy of spheres. However, we identified one parameter $\lh$ in the action \eqref{eq:GeneralAction2ndOrder} which sets the graviton normalization and turns out to be sensitive to more detailed bulk structures such as higher curvature terms. We will now argue that this parameter does not enter the second order entanglement entropy.\footnote{ See \S\ref{sec:1stLaw} for related observations based on the entanglement first law.} This is the gravity version of the statement that \eqref{eq:SEECFT} is independent of $\CT$.

We recall that due to the spherical symmetry of $\mathcal{A}$, extrinsic curvatures on the minimal surface $\mathcal{E}_{min}$ vanish and the entanglement entropy functional to be evaluated on $\mathcal{E}_{min}$ is just the Wald functional. In vacuum this reduces to just the area of $\mathcal{E}_{min}$ multiplied by a normalization which is basically determined by $a_d^*$ as in \eqref{eq:WaldEntropy}.
The perturbation can then be computed in holography as 
\begin{equation}\label{eq:DeltaWald}
 \delta S_{EE}(\mathcal{A}) = \delta S_{Wald} (\mathcal{E}_{min}) =  \delta \left( -2\pi \int_{\mathcal{E}_{min}} \sqrt{\gamma} \; \frac{\delta \mathcal{L}}{\delta R_{abcd}} \,n_{ab} \, n_{cd} \right)\,.
\end{equation}
Clearly $\delta S_{Wald}(\mathcal{E}_{min})$ can be expanded perturbatively by considering different powers of $h_{ab}$ and $\phi$. From the structure of the gravitational backreaction it follows that there is no contribution at $\order(\lambda)$. At $\order(\lambda^2)$ (i.e., at first order in $h_{ab}$) the variation in \eqref{eq:DeltaWald} acts on the objects inside the integral, but it does not change the surface $\mathcal{E}_{min}$ itself.\footnote{ Variations of the bifurcation surface only affect the Wald entropy at the next order, which is beyond our discussion. To see this, note that the surface $\mathcal{E}_{min}$ extremizes the Wald entropy functional upon variation with respect to the metric. First order variations of the metric (which are $\order(\lambda^2)$ in our conventions) hence leave $\mathcal{E}_{min}$ unchanged.}
Since the surface $\mathcal{E}_{min}$ is maximally symmetric, we anticipate that this variation of the Wald functional is just a variation of the area form (up to normalization). That is, the evaluation of \eqref{eq:DeltaWald} should yield precisely the same answer as in Einstein gravity up to normalization. Let us make this slightly more explicit to figure out the correct normalization.

To evaluate \eqref{eq:DeltaWald} at $\order(\lambda^2)$, we should expand the integrand linearly in the perturbation:
\begin{equation}\label{eq:DeltaWald2}
 \delta S_{Wald}(\mathcal{E}_{min}) =   -2\pi \int_{\mathcal{E}_{min}}  \left[\sqrt{\gamma} \; \frac{\delta \mathcal{L}}{\delta R_{abcd}} \,n_{ab} \, n_{cd} \right]_{\order(h)} \,.
\end{equation}
The obvious way to evaluate this is to start with the full non-linear Lagrangian, expand the Wald entropy density around the AdS background, and then identify the $\order(h)$ term in \eqref{eq:DeltaWald2}. However, even without knowledge of the full theory it is clear that ultimately the integrand can only be sensitive to the part of $\mathcal{L}$ which is at most quadratic in $h_{ab}$, i.e., we should be able to predict the answer just based on the most general second order Lagrangian \eqref{eq:GeneralAction2ndOrder}. Moreover, only the purely gravitational sector of \eqref{eq:GeneralAction2ndOrder} is relevant for Wald entropy at $\order(\lambda^2)$. This is quite clear from the observations that a minimally coupled scalar field in gravity does not couple to curvature and the action \eqref{eq:GeneralAction2ndOrder} agrees with that of a minimally coupled scalar at the order we are interested in. 

From these observations, we are led to conclude that the integrand in \eqref{eq:DeltaWald2} formally has the same structure in any allowed theory of gravity up to the normalization $\lh$. Therefore, without loss of generality, we can compute the integral as if it came from Einstein's theory and multiply by $\lh$. But in that case, the answer for $\delta S_{Wald}(\mathcal{E}_{min})$ is well known: it is just the variation of the area functional. Therefore, \eqref{eq:DeltaWald2} can be rewritten in general as a variation of area, multiplied by the normalization $\lh$ that encodes the higher derivative dependence:
\begin{equation}\label{eq:deltaSWald4}
\begin{split}
  \delta &S_{Wald}(\mathcal{E}_{min}) =
   2\pi\left(\frac{\lAdS^{d-1}}{\lP^{d-1}}\, \lh \right) \frac{\int_{\mathcal{E}_{min}} \sqrt{\bar{\gamma}} \; (\tfrac{1}{2}\bar\gamma^{\alpha\beta} h_{\alpha\beta})}{\lAdS^{d-1}} + \order(h^2) \,.
\end{split}
\end{equation}
where $\bar\gamma_{\alpha\beta}$ is the background metric induced on $\mathcal{E}_{min}$ by $\gn_{ab}$ and $h_{\alpha\beta}$ is the induced metric perturbation.\footnote{ Explicitly, if $\xi^\alpha(x)$ denotes the coordinates of $\mathcal{E}_{min}$ embedded in AdS${}_{d+1}$, we have
\begin{equation}
  \bar\gamma_{\alpha\beta} = \frac{\partial x^a}{\partial \xi^\alpha} \frac{\partial x^b}{\partial \xi^\beta} \, \gn_{ab} \,,\qquad
  h_{\alpha\beta} = \frac{\partial x^a}{\partial \xi^\alpha} \frac{\partial x^b}{\partial \xi^\beta} \, h_{ab} \,.
\end{equation}
For more conventions see \S\ref{app:Wald}.}
Note that the arguments just given were heuristic and we are discarding boundary terms which would be important to regulate the divergences. A detailed derivation of \eqref{eq:deltaSWald4} can be found in \S\ref{app:Wald}.

The integral \eqref{eq:deltaSWald4} can now be used to derive the independence of $\delta S_{Wald}(\mathcal{E}_{min})$ on $\lh$. We just need to find the on-shell dependence of $h_{\alpha\beta}$ on $\lh$ and combine it with the prefactor in \eqref{eq:deltaSWald4}. 
To this end, consider the linearized equations of motion implied by the general action \eqref{eq:GeneralAction2ndOrder}:
\begin{equation}\label{eq:generalEOM0}
\begin{split}
 \lh \, \prop_{ab}{}^{cd} \, h_{cd} = - \frac{\lphi}{2} \, T_{ab}^{(\phi)} \,.
\end{split}
\end{equation}
From this equation it is clear how to go from the universality of graviton propagation to the universality of perturbed Wald entropy: apart from the overall $\lh$ every dimensionless parameter in \eqref{eq:generalEOM0} has been fixed by the boundary input $\{a_d^*,\Delta,\cOO\}$. The combination $\lh h_{cd}$ therefore satisfies a universal equation of motion which is independent of the non-linear details of the bulk theory. Since the perturbed Wald entropy \eqref{eq:deltaSWald4} depends on precisely this particular combination it is clear that $\delta S_{Wald}(\mathcal{E}_{min})$ will be the same in all theories we consider. Said differently, the function $h_{ab}$ which satisfies the backreaction equation \eqref{eq:generalEOM0} is proportional to $1/\lh$ such that the $\lh$ dependence in \eqref{eq:deltaSWald4} cancels. We conclude that $\delta S_{EE}(\mathcal{A}) = \delta S_{Wald}(\mathcal{E}_{min})$ is independent of $\lh$. This is consistent with the field theory result \eqref{eq:SEECFT} being independent of $\CT$.  

This concludes our analysis of the role which various dimensionless parameters play.
Having verified the cancellation of $\lh$, the explicit solution of \eqref{eq:generalEOM0} and evaluation of \eqref{eq:deltaSWald4} can be done without loss of generality at $\lh = 1$, i.e., Einstein gravity. 
We refrain from repeating this calculation. See, for example, \cite{Liu:2012eea,Nishioka:2014kpa} for an explicit evaluation of \eqref{eq:deltaSWald4} in the case of a scalar field source. The result of their computation perfectly matches the CFT answer \eqref{eq:SEECFT}.

\section{Example: Gauss-Bonnet theory}
\label{sec:exampleGB}

In this section we wish to illustrate the general argument, using the example of gravity theories with Lagrangians quadratic in curvature.  This discussion will make contact with a number of previous higher derivative computations of entanglement entropy. In particular we will show that various results for perturbations of entanglement entropy of spherical regions dual to higher derivative theories can be equally well reproduced from a calculation in cosmological Einstein gravity by renormalizing the discrete set of free parameters which play a role at second order in perturbation theory, i.e., $\{\lAdS/\lP \,,\, m^2\lAdS^2 \,, \lphi\}$. We will also see the appearance of the additional parameter $\lh$ which drops out of the formula for entropy.

\subsection{Two gravity models}
\label{sec:models}

The basic gravitational model that we want to take as the reference system is just the auxiliary Einstein-Hilbert theory described by
    \begin{equation} \label{eq:EHmod}
    \begin{split}
      I = \frac{1}{2\lP^{d-1}}\int \sqrt{-g} \, \left[R + \frac{d(d-1)}{\ell^2}
          - \frac{\lphi}{2}\left(  (\nabla \phi)^2 + m^2 \, \phi^2 \right)\right] \,,
          \end{split}
          \end{equation}
       with asymptotic boundary condition ${\phi}(z\rightarrow 0) \simeq  \lambda\,z^{d-\Delta}$ such that ${\phi}$ sources a scalar operator with dimension $\Delta$ and coupling $\lambda$ in the CFT. We parametrize the lowest order backreaction of the scalar probe on the geometry as $g_{ab} = \gn_{ab} + h_{ab}$, where $\gn_{ab}$ is AdS${}_{d+1}$ with scale $\lAdS^2 = \ell^2$. The independent dimensionless parameters in the gravitational problem can therefore be taken as $\{(\lAdS/\lP)\,,\,m^2\lAdS^2\,,\,\lphi\}$, while $\lh = 1$ for Einstein gravity.

In this section we wish to reproduce predictions of four-derivative theories by using the above Einstein theory with renormalized parameters.
The only four-derivative theory which passes the consistency condition of yielding second order equations of motion is Gauss-Bonnet theory, which we will discuss here. Nevertheless, in \S\ref{app:example} we explore the most general four-derivative theory of gravity and come to very similar conclusions in the case where the higher derivative terms are small perturbative corrections to \eqref{eq:EHmod}. 
 
With this motivation in mind, let us take the following action as the theory to compare with:
      \begin{equation} \label{eq:HigherDerPert}
      \begin{split}
        \tilde{I} &= \frac{1}{2\lPt^{d-1}}\int  \sqrt{-g} \, \left[R + \frac{d(d-1)}{\ellt^2} + \ellt^2\,\mathcal{L}_{h.d.} 
            - \frac{\lphit}{2} \left( (\nabla  \phi)^2 + \tilde{m}^2 \, \phi^2 \right)\right] \\
            &\quad \text{where } \mathcal{L}_{h.d.} = \gamma  \left(R^2 - 4 R_{ab} R^{ab} + R_{abcd} R^{abcd}\right)
       \end{split}
       \end{equation}
again with a fixed boundary condition ${\phi}(z\rightarrow 0) \simeq  \lambda\,z^{d-\Delta}$. The AdS${}_{d+1}$ background solution of this theory will have some other scale $\lAdSt$ which is a function of $\ellt$ and $\gamma$. 
In the context of both theories $I$ and $\tilde{I}$, we will use $\gn_{ab}$ to refer to the AdS${}_{d+1}$ background. From the context it should be clear whether $\gn_{ab}$ is the background with $\lAdS$ or with $\lAdSt$.

Our basic aim is to show that for the purpose of carrying out the perturbed entanglement entropy computation of spherical regions, the actions \eqref{eq:EHmod} and \eqref{eq:HigherDerPert} are indistinguishable if the dimensionless parameters $\{(\lAdS/\lP)\,,\,m^2 \lAdS^2\,,\,\lphi\}$ are suitably expressed as functions of $\{(\lAdSt/\lPt)\,,\,\tilde{m}^2 \lAdSt^2\,,\,\lphit\,,\,\gamma\}$ such that they are consistent with the CFT boundary conditions. In order to show this, we will now derive the relevant ingredients for such a computation order by order in $\lambda$ from both $I$ and $\tilde I$. We will demonstrate that the parameters of Einstein gravity can always be chosen such that the two perturbative expansions match. The higher derivative coupling $\gamma$ can be absorbed entirely into renormalizations of the other dimensionless quantities.

\subsection{Parameter matching}
\label{sec:matching}

We will now determine the relation between the dimensionless parameters of the models $I$ and $\tilde I$ such that both make identical predictions for vacuum and perturbed entanglement entropy of spheres.

\paragraph{Finding the AdS${}_{d+1}$ background solution:}

Before turning on the scalar perturbation, we need to determine the maximally symmetric AdS${}_{d+1}$ background solutions of the two theories \eqref{eq:EHmod} and \eqref{eq:HigherDerPert}. It is well known that such solutions exist in both theories and are ghost-free \cite{Boulware:1985aa} if we choose the respective AdS scales
\begin{equation}\label{eq:BGsols}
\begin{split}
  \lAdS^2 &= \ell^2 \,,\\
  \lAdSt^2 &= \frac{\ellt^2}{2} \left( 1+ \sqrt{1- 4(d-2)(d-3) \, \gamma }\,\right) \,.
\end{split}
\end{equation}
We will henceforth expand around the AdS${}_{d+1}$ background and thus treat $\lAdS$ and $\ell$ as equivalent (and similarly with tildes).

\paragraph{Matching of Wald entropy (or $a_d^*$):}

Let us remind the reader of the result of \S\ref{sec:introduction} that Wald entropy of maximally symmetric spaces is always proportional to area. We can thus compute the unperturbed entanglement entropy of $\mathcal{A}$ as the Wald entropy of hyperbolic black holes corresponding to the two AdS radii \eqref{eq:BGsols} of Einstein and Gauss-Bonnet gravity, respectively. By integrating \eqref{eq:WaldEntropy} for the two gravity models, one finds
\begin{equation}
\begin{split}
 S_{Wald} &= \frac{2\pi}{\pi^{d/2}} \Gamma\left(d/2\right) \, a_d^* \; \frac{\text{Vol}(\mathbb{H}_{d-1})}{\lAdS^{d-1}}\,, \qquad
 a_d^* = \frac{\pi^{d/2}}{\Gamma(d/2)} \frac{\lAdS^{d-1}}{\lP^{d-1}} \,,\\
 \tilde{S}_{Wald} &= \frac{2\pi}{\pi^{d/2}} \Gamma\left(d/2\right) \, \tilde{a}_d^* \; \frac{\text{Vol}(\mathbb{H}_{d-1})}{\lAdSt^{d-1}} \,, \qquad
 \tilde{a}_d^* = \frac{\pi^{d/2}}{\Gamma(d/2)} \frac{\lAdSt^{d-1}}{\lPt^{d-1}} \left( 1 - 2(d-1)(d-2) \,\gamma\, \frac{\ellt^2}{\lAdSt^2}\right) \,.
\end{split}
\end{equation}
Given the entanglement entropy $\Sn_{EE}(\mathcal{A})$ with fixed normalization in the CFT, it can be reproduced by the computation of Wald entropy in either of the gravity models by requiring $a_d^* = \tilde{a}_d^*$, i.e.,
\begin{equation}\label{eq:lPrenorm}
\frac{\lAdS^{d-1}}{\lP^{d-1}} =\frac{\lAdSt^{d-1}}{\lPt^{d-1}} \left( 1 - 2(d-1)(d-2) \,\gamma\, \frac{\ellt^2}{\lAdSt^2}\right) \,.
\end{equation}
This determines the dimensionless ratio $\lAdS/\lP$ in Einstein gravity in terms of $\{\lAdSt/\lPt\,,\,\gamma\}$ in Gauss-Bonnet theory such that both theories make the same predictions for Wald entropy of the maximally symmetric horizon. 
From the CFT point of view, this matching is the condition that $a_d^*$ is a fixed fundamental parameter of the CFT which should be reproduced by either candidate of a gravity dual.

\paragraph{Matching of scalar coupling (or $\cOO$):}

Perturbing the CFT with $\mathcal{O}$ means introducing a massive scalar in the gravitational action. The mass of the scalar is constrained by the requirement that the boundary value of the scalar sources the operator $\mathcal{O}$ with given dimension $\Delta$. This fixes
\begin{equation}\label{eq:MassRel}
\lAdS^2\,m^2 = \Delta (\Delta-d) = \lAdSt^2\,\tilde{m}^2 \,.
\end{equation}
The dynamics of the scalar field is derived at lowest order from the propagation on AdS${}_{d+1}$ vacuum:
\begin{equation} \label{eq:ScalarSol}
\left(\Boxn - m^2\right)\phi = 0 \,,\qquad \phi(z \rightarrow 0) \sim \lambda \, z^{d-\Delta} \,,
\end{equation}
and similarly with tildes. Once the boundary condition is imposed in this way, standard holographic techniques lead to the matching between the normalization $\cOO$ of the relevant CFT correlators and the normalization of the scalar field action via \eqref{eq:cOOmatch}: the two gravity theories $I$ and $\tilde I$ lead to the same $\cOO$ provided that
\begin{equation}
 \lphi \, \frac{\lAdS^{d-1}}{\lP^{d-1}}= \lphit \,\frac{\lAdSt^{d-1}}{\lPt^{d-1}}  \,,
\end{equation}
so the ratio $\lphi/\lphit$ should be fixed to be the same as in \eqref{eq:lPrenorm}.

\paragraph{Computation of perturbed entanglement entropy:}

We finally want to check explicitly the arguments of \S\ref{sec:lhindependence} and verify that after the matching conditions above have been implemented, second order entanglement entropy is automatically the same for $I$ and $\tilde I$. 
We will also verify that the difference between $\lh=1$ and $\lht$ cannot be absorbed in any further free parameters. This will lead to the conclusion that the central charge $\CT$ is genuinely different in the CFTs dual to $I$ compared to those dual to $\tilde I$ after the above matchings have been performed.

Our perturbative scheme in the CFT demands that we consider the linearized equations of motion in the bulk. The linearized equations of motion can be obtained from the actions \eqref{eq:EHmod}, \eqref{eq:HigherDerPert} by linearizing around $\gn_{\mu\nu} = g_{\mu\nu} -  h_{\mu\nu}$. 
We find for $I$ and $\tilde I$, respectively \cite{Deser:2002jk}:
\begin{equation}\label{eq:GravEOM}
\begin{split}
 \lh \,\prop_{ab}{}^{cd} h_{cd}  &=  -\frac{\lphi}{2}\, T_{ab}^{(\phi)} \,, \\
 \lht\, \tilde{\prop}_{ab}{}^{cd} h_{cd}
  &= -\frac{\lphit}{2} \,  \, \tilde{T}_{ab}^{(\phi)} \,,\qquad
\end{split}
\end{equation}
where ${T}_{ab}^{(\phi)}$ and $\tilde{T}_{ab}^{(\phi)}$ are the source terms \eqref{eq:Tphi} for scalar matter with mass $m^2$ and $\tilde{m}^2$, respectively propagating on AdS${}_{d+1}$ with scale $\lAdS$ and $\lAdSt$. The kinetic operators $\prop_{abcd}$ and $\tilde\prop_{abcd}$ are defined as in \eqref{eq:EHlichnerowicz}, but, of course, also refer to backgrounds with respective scales $\lAdS$ and $\lAdSt$. The coefficients $\lh$ and $\lht$ in \eqref{eq:GravEOM} can be computed for $I$ and $\tilde I$ theories and turn out to be
\begin{equation}\label{eq:ratioR}
\lh = 1 \,,\qquad
 \lht = \left( 1 - 2(d-2)(d-3) \, \gamma \, \frac{\ellt^2}{\lAdSt^2}\right)\,.
\end{equation}
Clearly there is no free parameter left in \eqref{eq:GravEOM} which could be renormalized to absorb these normalizations. From \eqref{eq:cTmatch} it is hence clear that the $\CT$ charges of the dual CFTs will differ by the ratio $\lh/\lht$. At this stage, any CFT observable which did depend on $\CT$ would therefore genuinely distinguish between the two gravity models. As shown before, $\delta S_{EE}(\mathcal{A})$ does not depend on $\CT$, though.

Our matching prescriptions \eqref{eq:BGsols}, \eqref{eq:lPrenorm} and the graviton kinetic term scaling \eqref{eq:ratioR} have, of course, appeared in the literature before.  
For example, see \cite{deBoer:2011wk} for a computation of perturbed sphere entanglement entropy in Gauss-Bonnet theory. One can easily verify that their computation gives an answer which takes the form of perturbed entanglement entropy in Einstein gravity with dimensionless parameters rescaled as predicted by our general analysis.\footnote{ See, e.g., Eq.\ (7.7) of \cite{deBoer:2011wk} with the understanding that their Gauss-Bonnet coupling $\lambda$ equals $2\gamma$ in our notation.} At this order the entanglement entropy computation for spherical regions is therefore not sensitive to the Gauss-Bonnet coupling.

\section{Discussion}
\label{sec:discussion}

Let us summarize our arguments and comment on implications for the program of bulk reconstruction.

\paragraph{Summary:}
We considered the second order perturbation of CFT vacuum entanglement entropy of spheres due to scalar operator deformations. This quantity exhibits a universal functional form which only depends on a small number of fundamental CFT parameters to set various normalizations. We explored a similar universality statement about gravity, which explains why the corresponding holographic computation can always be recast in terms of one universal gravitational system, viz., linearized Einstein gravity coupled to a free massive scalar field. We argued that any ghost-free gravitational theory other than Einstein gravity can be used to make identical predictions due to the uniqueness of diffeomorphism invariant spin-2 propagation on AdS. 
For instance, we demonstrated how in Gauss-Bonnet theory both the vacuum and the perturbed sphere entanglement entropy (obtained by computing Wald entropy of a minimal surface) can be reproduced from just Einstein gravity by suitably renormalizing the dimensionless parameters of the theory. In this sense, knowledge of this entanglement data does not allow to distinguish between Gauss-Bonnet theory and Einstein gravity.

More precisely, the perturbed sphere entanglement only depends on the dimension $\Delta$ of the deforming operator $\mathcal{O}$, the generalized $a$-central charge $a_d^*$ (in vacuum) and the normalization $\cOO$ of the $\langle\mathcal{O}\mathcal{O}\rangle$ correlator (at second order in the perturbation). We saw that any viable gravity model has enough dimensionless parameters to account for these dependencies (the scalar mass $\lAdS^2m^2$, the normalization of the gravitational action, $\lAdS/\lP$, and the normalization of the scalar field, $\lphi$, respectively). In particular, no higher curvature terms in the bulk are required in order to obtain a sufficient number of couplings for consistently performing the holographic matching. In this problem, the only signature of higher curvature couplings would be a renormalization of the parameters of cosmological Einstein gravity. This implies that perturbed entanglement entropy of spheres can be computed using any ghost-free gravitational model as an auxiliary system, even for CFTs without any simple gravity dual at all \cite{Faulkner:2014jva}.


In face of this, it is clear that for the derivation of bulk dynamics from entanglement in the sense of AdS/CFT it is desirable to reconstruct bulk features which are not entirely universal in the above sense. 
The obvious question arises which quantities one should consider instead of or in addition to entanglement entropy of spheres in order to achieve a dependence on bulk features which are not the same for all CFTs. We will come back to this question below.

\paragraph{Generalizations:}
There are some modifications of our analysis which would be interesting to explore. 
For instance, one could compute the perturbation to higher order in $\lambda$. By extending the perturbation theory of \S\ref{sec:CFTExplanation}, it is clear that entanglement entropy will then be sensitive to higher point correlation functions which would probe bulk dynamics beyond the linear approximation. While this would be very interesting to see explicitly, the major complication in the bulk would be a backreaction not just on the Wald entropy density, but on the minimal surface itself. Once the minimal surface has non-vanishing extrinsic curvatures, ambiguities in the entanglement entropy functional need to be addressed \cite{Dong:2013qoa,Camps:2013zua,Bhattacharyya:2013gra,Bhattacharyya:2014yga}. By resolving this issue, one may hope to derive bulk dynamics beyond linear order.

We mainly focused on perturbations of entanglement due to deformations of the CFT by a relevant scalar operator. As another generalization, one can imagine deformations by operators of other spin. As long as the deformation does not involve a stress tensor, our conclusions are expected to hold since the new operator comes at least with an overall normalization which is all that is required to account for the renormalization of $\lAdS/\lP$. Consider, however, the case where the CFT is deformed by a stress tensor insertion. In that case, the first interesting perturbation of the entanglement entropy occurs at $\order(\lambda)$ instead of $\order(\lambda^2)$, where one gets a contribution $\sim \langle T T \rangle$ with normalization set by $\CT$. In the bulk the problem of gravitational backreaction changes in a similar way and the extremal surface will feel a perturbation at $\order(\lambda)$. It would be interesting to explore this case in detail and consider the dependence on $\CT$ to draw similar conclusions about the effect of higher curvature couplings. The dependence on $\CT$ is expected to lead to a way of distinguishing Einstein gravity duals from higher curvature ones. In fact it has been argued before (as we review in \S\ref{sec:1stLaw}) that generic perturbations of spherical region entanglement entropy can be used to derive linearized bulk equations including their normalization set by $\CT$ \cite{Faulkner:2013ica}.

Further generalizations could involve perturbations of the CFT other than operator deformations. For instance, a small geometric perturbation of the spatial sphere should lead to similar conclusions as those just mentioned. Similarly, $1/N$ corrections have recently been explored in order to improve on the entanglement first law \cite{Kelly:2015mna}. It would be interesting to pursue this further and explore ways to reconstruct bulk dynamics at a level where one can distinguish between different higher curvature theories. Instead of considering individual approaches in detail, let us now outline in general how a good choice of CFT observables can be made which are sensitive to interesting bulk features in a controlled way.

\paragraph{How to ``bootstrap'' bulk theories:}
As we have just seen, there are, of course, simple ways to circumvent the problem of perturbed sphere entanglement being insensitive to the desired detailed bulk information. Roughly speaking, one needs to consider more detailed CFT structures in order to distinguish between different bulk models. In general one could imagine bootstrapping the bulk theory using the CFT input. Clearly it is hard to derive bulk dynamics directly from fundamental CFT parameters such as two-point functions and OPE coefficients. But entanglement entropy seems to provide a physically meaningful quantity with well-understood universal properties on both sides of the duality, which depends on these fundamental building blocks of the CFT in a useful way. At lowest order, the vacuum entanglement entropy of spheres can be used to measure the generalized $a$-charge $a_d^*$. In our particular context we used this to fix the normalization $\lAdS/\lP$ of any possible bulk dual (whose dynamics is uniquely fixed in the same way as the functional form of entanglement entropy is universal). As we have seen, a small perturbation of this entanglement entropy does not affect the dynamics enough to determine if there are any higher curvature terms present in the bulk theory. To answer this question, one would have to measure some other entanglement entropy which is sensitive to further details of the bulk dynamics. 

One natural CFT parameter that we have only encountered briefly so far is the stress tensor central charge. Let us elaborate how any quantity which is sensitive to $\CT$ will also be sensitive to bulk higher derivative couplings and hence will not be reproducible by just Einstein gravity once $\lAdS/\lP$ has been fixed by matching vacuum entanglement entropy of spheres as outlined in this note. 
In our context, the ratio $a_d^*/\CT$ computed holographically is independent of $\lAdS/\lP$ and genuinely takes different values in the presence of higher curvature terms. In fact it is clear from \eqref{eq:adHD} and \eqref{eq:cTmatch} that higher curvature terms in the bulk are necessary whenever $a_d^* / \CT = 2 \pi^{d/2} /  \Gamma(d/2)f_d$ does {\it not} hold. 
This is, of course, just a generalization of the well-known statement that higher curvature terms in the bulk theory are required whenever the $a$- and $c$-central charges parametrizing the boundary conformal anomaly differ \cite{Henningson:1998gx,Blau:1999vz,Nojiri:1999mh}.

Any set of CFT observables which depends on both $a_d^*$ and $\CT$ can hence be used to detect if higher curvature terms in the bulk are necessary. The problem of genuine bulk reconstruction is therefore not at all the restriction to quantities which are universal in the sense that they only depend on an overall normalization. But it is now clear that one has to consider {\it enough} such quantities in order to see a necessity to account for their values in terms of higher curvature interactions. 

There are many examples of entanglement-related CFT quantities which one may envisage to compute in this context. For instance, the entanglement entropies of other surfaces such as infinite strips or cylinders are sensitive to both $a_d^*$ and $\CT$ \cite{Solodukhin:2008dh,Hung:2011xb,deBoer:2011wk}. Further, the presence of corners and kinks in the spatial region of interest leads to divergences which exhibit a universal dependence on the opening angle and (at least in holographic CFTs) depend on $a_d^*$ and $\CT$ in an interesting way \cite{Hirata:2006jx,Fradkin:2006,Casini:2006hu,Myers:2012vs,Bueno:2015rda,Bueno:2015xda,Alishahiha:2015goa,Miao:2015dua}. 
Other interesting quantities with universal features and dependence on $\CT$ are R\'enyi entropies of spherical regions (e.g., in the entanglement entropy limit $q\rightarrow 1$) \cite{Hung:2011nu,Galante:2013wta,Perlmutter:2013gua}.
All these quantities have in common a contribution with a universal functional form and a dependence on few CFT parameters (in particular $a_d^*$ and $\CT$). They could hence be used to distinguish between CFTs with a dual description in terms of Einstein gravity and CFTs which require higher curvature terms in the bulk.

However, note that even if some CFT measurement of $a_d^*$ and $\CT$ leads to the conclusion that there can exist a semiclassical bulk theory only if contains higher curvature interactions are present, it is usually not possible to determine precisely what higher curvature terms are required. Indeed, the signature of, say, a Gauss-Bonnet term on $a_d^*/\CT$ can equally well be reproduced by some other higher derivative term. For illustration of this statement see also \S\ref{app:Wald}, where it is shown how any higher curvature theory simplifies to the background field expansion of Gauss-Bonnet form if one is only concerned with linearized equations of motion. In this sense Gauss-Bonnet theory is the most general theory which can account for arbitrary values of $a_d^*$ and $\CT$ in the CFT. In order to further bootstrap the precise bulk dynamics, one needs again access to more CFT data (e.g., entanglement data which depends on further 2-point functions or OPE coefficients). The reason is, again, that gravitons propagating on AdS have very universal dynamics. This is a manifestation of the familiar problem that entanglement entropies which are computable tend to be very universal and contain only discrete information about the actual bulk dynamics. 

It would be interesting to use this philosophy to derive more detailed (non-linear) bulk dynamics for some simple CFT with known holographic dual from first principles. Generalizing this approach may then ultimately lead to a genuine reconstruction of the bulk geometry and its dynamics by using a suitable set of entanglement entropies.

\acknowledgments
I am very grateful to Jyotirmoy Bhattacharya and Mukund Rangamani for collaboration during early stages of this project, numerous insightful discussions and comments on a draft. I also wish to thank Rob Myers for useful comments.
I am supported by a Durham Doctoral Fellowship.

\appendix

\section{Generalization to bulk theories with higher order equations of motion}
\label{app:generalization}

In the main text, we considered the case where the equations of motion for the metric perturbation in the bulk were strictly second order. The motivation for doing so was to exclude ghost modes which would occur if the graviton equations were higher than second order. However, if the higher curvature contributions to gravity are parametrically small, they can be interpreted, e.g., as toy models for perturbative string theory corrections to Einstein gravity. The ghost modes are then an artefact of the low energy truncation and can be ignored, which allows for a considerable increases of the number of acceptable gravitational models covered by our analysis. In order to illustrate this, we first generalize in \S\ref{app:example} the Gauss-Bonnet theory of \S\ref{sec:exampleGB} to the case of all Lagrangians which are quadratic in curvature. In \S\ref{app:generalArgument} we then give a more abstract argument to show that this example actually illustrates the general case.

\subsection{Example: general curvature squared theories}
\label{app:example}

In this section we generalize the analysis of \S\ref{sec:exampleGB} to theories where the gravitational sector is described by 
 \begin{equation} \label{eq:HigherDerPert2}
 \begin{split}
 \tilde{I} &= \frac{1}{2\lPt^{d-1}}\int  \sqrt{-g} \, \left[R + \frac{d(d-1)}{\ellt^2} + \ellt^2\,\mathcal{L}_{h.d.} 
            - \frac{\lphit}{2} \left( (\nabla  \phi)^2 + \tilde{m}^2 \, \phi^2 \right)\right] \\
            &\quad \text{where } \mathcal{L}_{h.d.} = \alpha \, R^2 + \beta\, R_{ab} R^{ab} +\gamma  \left(R^2 - 4 R_{ab} R^{ab} + R_{abcd} R^{abcd}\right)
 \end{split}
 \end{equation}
We do this to explicitly demonstrate that such theories -- despite the presence of massive ghost modes -- are acceptable in a perturbative sense. Indeed, we will see how the ghost degrees of freedom are suppressed and hence can be neglected for the computations we want to perform.

We want to model the theory \eqref{eq:HigherDerPert2} with the cosmological Einstein-scalar theory \eqref{eq:EHmod}. That is, we will give the matching conditions which express the parameters of the latter theory, $\{(\lAdS/\lP)\,,m^2\lAdS^2\,,\lphi\}$, in terms of $\{(\lAdSt/\lPt)\,,\,\tilde{m}^2 \lAdSt^2\,,\,\lht\,,\,\lphit,\alpha,\beta,\gamma\}$ such that the theories make the same predictions for entanglement entropy of spheres in vacuum even after deforming the theory with a scalar operator. The main new ingredient is the presence of four-derivative terms in the equations of motion. 

Before discussing the linearized equations of motion, let us give the analogue of the matching conditions \eqref{eq:BGsols}, \eqref{eq:lPrenorm}.
The stable AdS${}_{d+1}$ background solutions in the two theories have the following scales in terms of the cosmological constant:
\begin{equation}\label{eq:BGsols2}
\begin{split}
  \lAdS^2 &= \ell^2 \,,\\
  \lAdSt^2 &= \frac{\ellt^2}{2} \left( 1+ \sqrt{1- \frac{4(d-3)}{(d-1)} \, \big(d(d+1)\,\alpha+d\,\beta + (d-1)(d-2)\, \gamma\big) }\,\right) \,.
\end{split}
\end{equation}
The generalization of the $a_d^*$ matching condition \eqref{eq:lPrenorm} then becomes 
\begin{equation}\label{eq:lPrenorm2}
\frac{\lAdS^{d-1}}{\lP^{d-1}} =\frac{\lAdSt^{d-1}}{\lPt^{d-1}} \left( 1 - 2\big(d (d+1)\alpha+d\,\beta + (d-1)(d-2) \,\gamma\big) \frac{\ellt^2}{\lAdSt^2}\right) \,.
\end{equation}
Exactly as in \S\ref{sec:matching}, the previous equation also immediately fixes the ratio of scalar couplings, i.e., $\lphit/\lphi$ is given by the value bracket on the right hand side of \eqref{eq:lPrenorm2}. This guarantees that holographic computations of $\langle \mathcal{O}\mathcal{O}\rangle$ yield a unique normalization $\cOO$. 

Consider now the linearized graviton equations in the two theories. For this, we find it convenient to gauge fix the metric. In the transverse traceless gauge ($\dn_a h^{ab} = 0 = h$) the universal linearized equations of motion \eqref{eq:GravEOM} now generalize to the following two equations for Einstein and curvature squared theories, respectively \cite{Deser:2002jk}:
\begin{equation}\label{eq:GravEOM2}
\begin{split}
 \lh \left( \Boxn + \frac{2}{\lAdS^2}\right) h_{ab}  &=  -\lphi\, T_{ab}^{(\phi)} \,, \\
 \lht\,\left( \Boxn + \frac{2}{\lAdSt^2}\right) h_{ab}
 +\beta \, \ell^2 \, \left( \Boxn + \frac{2}{\lAdSt^2}\right)^2 h_{ab} &= -\lphit \,  \, \tilde{T}_{ab}^{(\phi)} \,,\qquad
\end{split}
\end{equation}
where the coefficients $\lh$ and $\lht$ in \eqref{eq:GravEOM2} are now given as
\begin{equation}\label{eq:ratioR2}
\lh = 1 \,,\qquad
 \lht = \left( 1 - 2\big(d(d+1)\alpha+d\,\beta + (d-2)(d-3) \, \gamma \big) \frac{\ellt^2}{\lAdSt^2}\right)\,.
\end{equation}
We now want to argue that the second term in the second line of \eqref{eq:GravEOM2} can be neglected for our computation of entanglement entropy. Indeed, this term is obviously suppressed if $\beta$ is treated as a small parameter and in the physical on-shell limit $(\Boxn + 2/\lAdSt^2) h_{ab} \rightarrow 0$ (see also \cite{Bueno:2015xda}). Further, we can also rewrite the second line of \eqref{eq:GravEOM2} in the following factorized form: 
\begin{equation}\label{eq:mostGeneralEOM}
\begin{split}
 \beta \, \ell^2 \left(\Boxn + \frac{2}{\lAdSt^2}\right) \left( \Boxn + \frac{2}{\lAdSt^2} + \frac{\lht}{\beta \, \ell^2} \right) h_{ab} = -\lphit \, T^{(\phi)}_{ab} \,.
\end{split}
\end{equation} 
From this expression one can anticipate the two poles of the graviton propagator. The Fierz-Pauli type mass term $\sim 2/\lAdSt^2$ corresponds to the physical graviton mode. The other mass term $\sim (2/\lAdSt^2 + \lht/\beta \ell^2)$ schematically leads to the ghost which is very heavy for small $\beta$ and hence does not go on-shell at low energies. It can be seen as an artefact of truncating some well-behaved theory at four-derivative order.

\subsection{General argument}
\label{app:generalArgument}

Despite the previous subsection appearing to be restricted to the specific case of curvature squared theories, the conclusion is actually more general. If higher curvature terms, which appear in the gravitational action with some dimensionless coupling $\alpha$, lead to ghosts modes, then the latter generically have a mass which scales with $1/\alpha$. If these modes are interpreted as due to consistently truncating perturbative quantum gravity (i.e., $|\alpha|\ll 1$), the ghosts are unphysical and indeed never go on-shell at low energies. In such a perturbative framework the presence of ghosts is hence not due to an instability of the theory (the underlying full theory is assumed to be unitary), but it is just a rather harmless consequence of neglecting certain sectors of the theory which are not being probed at low energies.

To illustrate these statements, consider, for example, a gravity sector of the form $\mathcal{L}_{grav} = \mathcal{L}_{grav}(g_{ab},R_{abcd})$ such that the equations of motion are fourth order in derivatives. The most general four-derivative gravitational equations of motion linearized around AdS are precisely the same as those captured by the curvature squared theories in \eqref{eq:HigherDerPert2}. This can be shown using the method of background field expansion; see, e.g., \cite{Amsel:2012se}, where it was indeed argued that the linearized action of any higher derivative theory of this type is the same as that of the most general curvature squared theory.\footnote{ In this context, see also \S\ref{app:Wald}.} Concretely, given some such higher curvature theory, the action expanded to second order around AdS takes the form \eqref{eq:HigherDerPert2} with parameters $\{\alpha,\beta,\gamma\}$ determined as linear combinations of the original higher derivative couplings. We can therefore ignore the ghosts of these theories as long as the higher derivative couplings are viewed as a semiclassical truncation stemming from some unitary UV-complete theory.

\section{Perturbations of Wald entropy: background field expansion}
\label{app:Wald}

This appendix complements the analysis of \S\ref{sec:lhindependence} by computing the expansion of the Wald entropy density at $\order(h_{ab})$, derived from the general second order Lagrangian \eqref{eq:GeneralAction2ndOrder}. This analysis will also serve to illuminate how the universality of graviton propagation on AdS simplifies the method of background field expansion, thus demonstrating a more general universality statement about quantities which depend on both $a_d^*$ and $\CT$. Indeed, we will show in what sense Einstein-Gauss-Bonnet theory is the most general theory at $\order(h^2)$. Any other physically acceptable higher curvature theory has the same background field expansion at this order.

It is straightforward that we can focus on the pure gravity part of the general Lagrangian \eqref{eq:GeneralAction2ndOrder} since the minimally coupled matter sector does not couple to curvature. We wish to give a similar argument for the constant term.  
Ignoring the matter sector puts us in the realm of \cite{Sen:2014nfa} (see also \cite{Faulkner:2013ica}), where a detailed analysis of perturbed Wald entropy on a maximally symmetric AdS background has been performed. The best strategy to compute the perturbative expansion of Wald entropy is to rewrite the expansion in $h_{ab}$ as a background field expansion of curvatures $R_{abcd}$ around $-\frac{1}{\lAdS^2}(g_{ac}g_{bd}-g_{ad}g_{bc})$. More explicitly, instead of writing explicit metric perturbations $h_{ab}$ we write the Lagrangian solely in terms of
\begin{equation}
\begin{split}
\Delta R_{abcd} &= R_{abcd} - \left[- \frac{1}{\lAdS^2} (g_{ac}g_{bd} - g_{ad} g_{bc})\right] \\
&  = \Ro_{abcd} + \frac{1}{\lAdS^2} \left( \gn_{ac} h_{bd} + \gn_{bd} h_{ac} - \gn_{ad}h_{bc} - \gn_{bc} h_{ad}\right)  + \mathcal{O}(h^2)\,,
\end{split}
\end{equation}
where $g_{ab}=\gn_{ab} + h_{ab}$ is the full metric and indices of expressions like this are raised with the full metric. 
This way, the pure gravity part of our general quadratic Lagrangian \eqref{eq:GeneralAction2ndOrder} can be interpreted as an expansion in $\Delta R_{abcd}$ and its contractions and derivatives. To wit, we can always write the pure gravitational couplings of \eqref{eq:GeneralAction2ndOrder} as a linear combination of all possible combinations of $\Delta R_{abcd}$ which contain terms relevant at $\order(h^2)$. Clearly such combinations contain at most two factors of $\Delta R_{abcd}$, but they may contain any number of derivatives in general:
\begin{equation}\label{eq:Rexpansion}
\begin{split}
  I_{grav} &= \frac{1}{2\lP^{d-1}} \int \sqrt{-\gn} \; \left(-\frac{2d}{\lAdS^2}\frac{\Gamma(d/2)}{\pi^{d/2}}\frac{\lP^{d-1}}{\lAdS^{d-1}} \, a_d^* + \frac{\lh}{2}\, h_{ab}\, \prop^{abcd}\,h_{cd} + \order(h^3)\right)  \\
           &= \frac{1}{2\lP^{d-1}} \int \sqrt{-g} \; \left( \mathcal{L}^{(I)}_{\Delta R} + \mathcal{L}^{(II)}_{\Delta R} \right) \,,\\
     \text{with } & \; \begin{cases}  \mathcal{L}^{(I)}_{\Delta R} \equiv a_0 +  a_2\, \Delta R \\
                   \mathcal{L}^{(II)}_{\Delta R} \equiv b_1 \,\lAdS^2\, (\Delta R)^2 + b_2 \, \lAdS^2\,\Delta R^{ab} \Delta R_{ab} + b_3 \,\lAdS^2\, \Delta R^{abcd} \Delta R_{abcd} +   \ldots + \order(h^3)  \end{cases}
\end{split}
\end{equation} 
Let us explain this way of parameterizing the action piece by piece. First observe that from the form of \eqref{eq:Rexpansion} it follows that $a_0$ is just 
\begin{equation}\label{eq:a0match}
a_0 =-\frac{2d}{\lAdS^2}\frac{\Gamma(d/2)}{\pi^{d/2}}\frac{\lP^{d-1}}{\lAdS^{d-1}} \, a_d^* \,.
\end{equation}
Similarly, one can easily verify that the vanishing of the linearized Lagrangian (i.e.\ the equation of motion for the background solution $\gn_{ab}$) demands $a_0 = -(2d/\lAdS^2)a_2$. This fixes the form of $\mathcal{L}^{(I)}_{\Delta R}$ and shows that the latter is all that would ever show up in the case of Einstein gravity (with an overall pre-factor $a_2$). 
Therefore, $\mathcal{L}^{(II)}_{\Delta R}$ is the term which parameterizes contributions to the graviton propagator which only come from higher curvature terms. More tensor structures involving derivatives and contractions of $\Delta R_{abcd}\Delta R_{efgh}$ could be written in \eqref{eq:Rexpansion} in order to match the expansion of an arbitrary theory of gravity at higher orders in $h_{ab}$. However, this is not needed to reproduce the first line, i.e., the universal parts up to $\order(h^2)$. To wit, the two lines of \eqref{eq:Rexpansion} match if we choose
\begin{equation}
  b_1 = -\frac{b_2}{4}  = b_3 = \frac{1}{4(d-2)} \,\left(\lh-\frac{\Gamma(d/2)}{\pi^{d/2}}\frac{\lP^{d-1}}{\lAdS^{d-1}} \, a_d^*\right) \,.
\end{equation}
Note that this choice is schematically of the Gauss-Bonnet form. We can interpret this as the statement that a Gauss-Bonnet term is all that is needed to reproduce the linearized graviton action with arbitrary graviton normalization. This is a generalization of the results of the main text: in CFT language, a quantity which is sensitive to both $a_d^*$ and $\CT$ may undeniably detect the presence of higher curvature terms, but it does not distinguish between Gauss-Bonnet theory and any other higher curvature theory in the same way that quantities only knowing about $a_d^*$ can always be computed using Einstein gravity. The reason is that linearized actions up to $\order(h^2)$ only require a small number of terms in the background field expansion.

As a consistency check, we can make the relation between parameters $\{a_0,a_2,b_1,b_2,b_3\}$ and $\CT$ manifest by referring to the results of \cite{Sen:2014nfa} (see also \cite{Faulkner:2013ica}). There it was shown that an action of the form \eqref{eq:Rexpansion} leads to $\lh = a_2 + 4(d-2) b_3$ which is proportional to $\CT$ via \eqref{eq:cTmatch}. This is in perfect agreement with the above statements.

Since \eqref{eq:Rexpansion} is manifestly written as both a perturbative expansion in the metric fluctuation and as a functional of curvatures, we can easily compute the associated perturbative expansion of Wald entropy density by taking derivatives with respect to $R_{abcd}$. Note that the way the expansion is written makes it explicit that any variation with respect to $R_{abcd}$ reduces the order in $h_{ab}$ by precisely one and the number of derivatives by two. The unperturbed Wald entropy is hence only sensitive to $\mathcal{L}^{(I)}_{\Delta R}$ and is readily checked to agree with \eqref{eq:WaldEntropy}:
\begin{equation}\label{eq:Wald00}
\begin{split}
  \Sn_{Wald}(\mathcal{E}_{min}) = -2\pi  \, \frac{1}{2\lP^{d-1}} \int_{\mathcal{E}_{min}} \sqrt{\bar\gamma} \, \left[\frac{\delta \mathcal{L}^{(I)}_{\Delta R}}{\delta R_{abcd}}\, n_{ab} n_{cd}\right]_{\text{AdS}} = \frac{2\pi}{\pi^{d/2}} \Gamma\left(d/2\right) \, a_d^* \, \frac{\int_{\mathcal{E}_{min}} \sqrt{\bar{\gamma}}}{\lAdS^{d-1}}\,,
\end{split}
\end{equation}
where the subscript `AdS' refers to evaluation on the background $\gn_{ab}$. 

Similarly, the perturbation of Wald entropy is now computable via the two contributions
\begin{equation}\label{eq:Pert0}
\begin{split}
  \delta S_{Wald}(\mathcal{E}_{min}) 
  = -2\pi  \, \frac{1}{2\lP^{d-1}} \int_{\mathcal{E}_{min}} \left[\sqrt{\gamma} \;\frac{\delta (\mathcal{L}^{(I)}_{\Delta R}+ \mathcal{L}^{(II)}_{\Delta R})}{\delta R_{abcd}}\, n_{ab} n_{cd}\right]_{\order(h)} \,,
\end{split}
\end{equation}
where `$\order(h)$' picks out the term proportional to $h_{ab}$. We find the following contribution from $\mathcal{L}^{(I)}_{\Delta R}$: 
\begin{equation}\label{eq:PertI}
\begin{split}
  \delta S_{Wald,\,(I)}(\mathcal{E}_{min}) 
  =  -2\pi  \, \frac{1}{2\lP^{d-1}}   \int_{\mathcal{E}_{min}} \left[\sqrt{{\gamma}}\;(-2a_2)\right]_{\order(h)}
  = \frac{2\pi}{\pi^{d/2}} \Gamma\left(d/2\right) \, a_d^* \, \frac{\int_{\mathcal{E}_{min}} \sqrt{\bar{\gamma}} \; (\tfrac{1}{2}\bar\gamma^{\alpha\beta} h_{\alpha\beta})}{\lAdS^{d-1}} \,,
\end{split}
\end{equation}
which could also be obtained by simply perturbing the volume form in \eqref{eq:Wald00}. 
Similarly, one can compute the higher curvature contributions from $\mathcal{L}^{(II)}_{\Delta R}$ at $\order(h)$ as follows:\footnote{ Our conventions are as follows: the embedded version of the intrinsic metric $\gamma_{\alpha\beta}$ is $\gamma_{ab} = g_{ab} + t_at_b-n_an_b$ where $t^a$ and $n^a$ are normal to the minimal surface such that $t^2 = -1$, $n^2 = 1$, $t^a n_a = 0$ and the binormal $n^{ab} \equiv n^{a} t^{b} - t^a n^b$.}
\begin{equation}\label{eq:PertII}
\begin{split}
  \delta &S_{Wald,\,(II)}(\mathcal{E}_{min}) =\\ 
  &\quad=  -2\pi  \, \frac{1}{2\lP^{d-1}}   \int_{\mathcal{E}_{min}} \sqrt{\bar{\gamma}} \;  2b_1\,\lAdS^2\bigg[ \Delta R \, g^{c[a} g^{b]d} - 4 \Delta R_{ef} \, g^{e[b} g^{a][c} g^{d]f} + \Delta R^{abcd} \bigg] n_{ab}n_{cd}  \\
  &\quad=  4\pi \,b_1\, \lAdS^2\, \frac{1}{\lP^{d-1}}  \int_{\mathcal{E}_{min}} \sqrt{\bar{\gamma}} \,\left[ R + 2 \, R^{ab} \, n_{ac} n_b{}^c - \frac{1}{2} \, R^{abcd} \, n_{ab} n_{cd}  \right]_{\order(h)}  + \order(h^2)  \\
  &\quad= 4\pi \,b_1\, \lAdS^2\, \frac{1}{\lP^{d-1}} \int_{\mathcal{E}_{min}} \sqrt{\bar{\gamma}} \; \mathcal{\Ro}+ \order(h^2) \,,
\end{split}
\end{equation}
where ${\mathcal \Ro}$ is the linearized intrinsic Ricci scalar on $\mathcal{E}_{min}$ in units of $\lAdS$ (i.e., \eqref{eq:PertII} is essentially the first order perturbation of Wald entropy in pure Gauss-Bonnet theory). In the last step we used the fact that extrinsic curvatures vanish on $\mathcal{E}_{min}$, so the projected Ricci scalar in the penultimate line coincides with the intrinsic one.
We can further simplify the expression using 
\begin{equation}
\mathcal{ \Ro} = - \bar{\mathcal{R}}^{\alpha\beta} \, \delta \gamma_{\alpha\beta} + \bar\nabla_\alpha (\ldots) = \frac{(d-2)}{\lAdS^2} \, \bar\gamma^{\alpha\beta} h_{\alpha\beta}+ \bar\nabla_\alpha (\ldots) \,,
\end{equation}
where $\bar{\mathcal{R}}^{\alpha\beta}$ is the unperturbed intrinsic Ricci scalar and $\bar \nabla_\alpha$ is the intrinsic covariant derivative compatible with $\bar\gamma_{\alpha\beta}$ and hence only yields boundary terms which we are not keeping track of.\footnote{ Boundary terms are important to get a finite result as in \eqref{eq:SEECFT}. Since we are only interested in identifying the fate of the dimensionless parameters, we are being implicit about these subtleties and assume that there always exist suitable Gibbons-Hawking-York boundary terms.} Hence \eqref{eq:PertII} becomes
\begin{equation}\label{eq:PertFinalII}
\begin{split}
  \delta &S_{Wald,\,(II)}(\mathcal{E}_{min}) =
   2\pi \left(\lh-\frac{\Gamma(d/2)}{\pi^{d/2}}\frac{\lP^{d-1}}{\lAdS^{d-1}} \, a_d^*\right) \frac{\int_{\mathcal{E}_{min}} \sqrt{\bar{\gamma}} \; (\tfrac{1}{2}\bar\gamma^{\alpha\beta} h_{\alpha\beta})}{\lP^{d-1}}
\end{split}
\end{equation}
Finally adding the two contributions to perturbed Wald entropy, \eqref{eq:PertI} and \eqref{eq:PertFinalII}, we conclude that \eqref{eq:Pert0} reads
\begin{equation}
\begin{split}
  \delta &S_{Wald}(\mathcal{E}_{min}) =
   2\pi\left(\frac{\lAdS^{d-1}}{\lP^{d-1}}\, \lh \right) \frac{\int_{\mathcal{E}_{min}} \sqrt{\bar{\gamma}} \; (\tfrac{1}{2}\bar\gamma^{\alpha\beta} h_{\alpha\beta})}{\lAdS^{d-1}}  \,.
\end{split}
\end{equation}
%


\section{First law of entanglement}
\label{sec:1stLaw}

In the main text we have focused on perturbations of entanglement entropy due to scalar deformations. Let us now drop such particular assumptions and consider a slightly more general setup which has been argued to serve as a quantitative derivation of bulk dynamics from entanglement. For theories in which the Ryu-Takayanagi proposal is valid, a first law of entanglement entropy has been shown to be equivalent to linearized bulk equations of motion \cite{Lashkari:2013koa,Blanco:2013joa,Wong:2013gua,Allahbakhshi:2013rda,Faulkner:2013ica,Swingle:2014uza}.  

The entanglement first law 
\begin{equation}\label{eq:EE1stLaw}
\delta S_{EE}(\mathcal{A}) = \text{Tr}(\delta \rho_\mathcal{A}\, H_\mathcal{A}) \equiv \delta \langle H_\mathcal{A} \rangle
\end{equation}
is purely a CFT statement which says that any change of $S_{EE}(\mathcal{A})$ due to some small perturbation $\delta \rho_\mathcal{A}$ of the density matrix is given by the change of modular energy associated with the spatial sphere $\mathcal{A}$. (This statement can easily be derived from the formalism of \S\ref{sec:CFTExplanation}.)
For CFTs with a semiclassical gravity dual, this statement has a natural analog in gravity: the left hand side of \eqref{eq:EE1stLaw} corresponds to a perturbation of Wald entropy as in \eqref{eq:DeltaWald}. Similarly the right hand side of \eqref{eq:EE1stLaw} translates to a change of gravitational energy $E_{grav}(\mathcal{A})$ of the Rindler patch of AdS${}_{d+1}$ associated with the causal diamond of $\mathcal{A}$.
That is, the CFT entanglement first law \eqref{eq:EE1stLaw} is equivalent to the gravitational first law
\begin{equation} \label{eq:Grav1stLaw}
\delta S_{Wald}(\mathcal{E}_{min}) = \delta E_{grav}(\mathcal{A}) \,.
\end{equation}
The gravitational energy $E_{grav}(\mathcal{A})$ is defined as an integral of the holographic stress tensor $T_{ab}^{grav}$ over a spatial bulk surface of bulk codimension two, which lies inside the Rindler patch of $\mathcal{A}$ and ends on $\partial \mathcal{A}$. Without loss of generality, we push this bulk surface to the boundary, so the gravitational energy can be computed as a boundary integral over $\mathcal{A}$: 
\begin{equation}\label{eq:ModEgrav}
 E_{grav}(\mathcal{A}) =  \int_{\mathcal{A}} d\Sigma^\mu \;  T^{grav}_{\mu\nu} \, \zeta^\nu_\mathcal{A}\,, 
\end{equation} 
where $d\Sigma^\mu$ is the volume form on the $(d-1)$-dimensional surface $\mathcal{A}$ and $\zeta^\nu_\mathcal{A}$ is the conformal Killing vector generating the Rindler flow inside the causal development of $\mathcal{A}$. In holographic CFTs the CFT stress tensor $T_{\mu\nu}$ of \eqref{eq:CFTstress} agrees with the holographic stress tensor $T_{\mu\nu}^{grav}$. One can easily verify that the gravitational energy \eqref{eq:ModEgrav} then agrees with the definition of modular energy in the CFT, \eqref{eq:ModHamDef}.

The holographic stress tensor is determined by the dominant asymptotics of the graviton fluctuation on AdS${}_{d+1}$. Let us work in Poincar\'e coordinates 
\begin{equation}\label{eq:coords}
 ds^2 = \frac{\lAdS^2}{z^2} \left( dz^2 +  \eta_{\mu\nu} \, dx^\mu dx^\nu \right) + h_{ab}\, dx^a dx^b\,.
\end{equation}
Then the relevant graviton falloff near the boundary is given by the following piece of the Fefferman-Graham expansion:
\begin{equation}
h_{ab}(x^\rho,z) \, dx^a dx^b = \lAdS^2 \, z^{d-2} \, h_{\mu\nu}^{(d)}(x^\rho) \, dx^\mu dx^\nu +\ldots \,. 
\end{equation}
The power of $z$ is chosen such that the induced boundary stress tensor $\delta T_{\mu\nu}^{grav}$ is finite (i.e., $z$-independent).
It was shown in \cite{Faulkner:2013ica,Sen:2014nfa} that for a wide class of theories the holographic stress tensor associated with this bulk perturbation evaluates to
\begin{equation}\label{eq:Tgrav}
 \delta T_{\mu\nu}^{grav} = \frac{d\,\CT}{f_d} \; h^{(d)}_{\mu\nu} \,,
\end{equation} 
where $f_d$ is the number defined in \eqref{eq:cTmatch}. Note that this form of $\delta T_{\mu\nu}^{grav}$ is consistent with our considerations in \S\ref{sec:lhindependence}. Indeed, $\delta S_{Wald}(\mathcal{E}_{min})$ in the present context can be obtained from the gravitational energy integral over the combination \eqref{eq:Tgrav}:
\begin{equation}
\delta S_{Wald}(\mathcal{E}_{min}) =  \frac{d\,\CT}{f_d}\int_{\mathcal{A}} d\Sigma^\mu \;  h^{(d)}_{\mu\nu} \, \zeta^\nu_\mathcal{A}\,.
\end{equation}
This has the same structure as \eqref{eq:deltaSWald4}, i.e., an integral over a quantity ($\CT$ or $\lh$ times a metric perturbation) which satisfies a universal equation of motion. 

The linearized equations of motion can be derived from the first law statements above. Indeed, assuming the Ryu-Takayanagi conjecture, it was shown in \cite{Faulkner:2013ica,Sen:2014nfa} that \eqref{eq:Grav1stLaw} for small spherical regions $\mathcal{A}$ is equivalent to linearized gravitational equations of motion for any higher derivative Lagrangian $\mathcal{L}_{grav} = \mathcal{L}_{grav}(g_{ab},R_{abcd},\nabla_eR_{abcd},\ldots)$. As argued in \S\ref{sec:GravExplanation}, on an AdS background these equations are the same for any theory of gravity with second order equations of motion, up to overall normalizations which encode the higher curvature contributions. 
Depending on the precise setup, even these normalizations may be absorbed into renormalizations of dimensionless quantities. In particular the normalization $\lh$ inside $\CT$ always multiplies $(\lAdS/\lP)^{d-1}$, c.f., \eqref{eq:cTmatch}. Unless $\lAdS/\lP$ has already been fixed (e.g., by a condition such as \eqref{eq:adHD}), this ratio can serve to account for any $\lh$ dependence.

\section{Second order gravitational perturbation theory}
\label{sec:perturbation}

In this appendix we list some relevant formulae for gravitational perturbation theory up to second order in the graviton. We take the ansatz
\begin{equation}
 g_{ab} = \gn_{ab} +  h_{ab}
\end{equation}
and are interested in perturbations up to $\order(h^2)$. The background solution $\gn_{ab}$ is AdS${}_{d+1}$ with curvature $-1/\lAdS^{2}$, so the following simplifications apply for the background curvature tensors:
\begin{equation} \label{eq:AdSsimplify}
\begin{split}
 \Rn_{abcd} &= -\frac{1}{\lAdS^2} (\gn_{ac} \gn_{bd} - \gn_{ad} \gn_{bc}) \,, \qquad
 \Rn_{ab} = - \frac{d}{\lAdS^2} \, \gn_{ab} \,, \qquad
 \Rn = - \frac{d(d+1)}{\lAdS^2}\,.
\end{split}
\end{equation}

Let us evaluate and define various quantities, where superscript numbers refer to the order in perturbation theory:
\begin{equation}
\begin{split}
  g^{ab} &= \gn{}^{ab} -  h^{ab} +  h^{ac}h_c^b + \order(h^3) \,,\\
  \Gamma^\rho_{ab} &= \Gamman^\rho_{ab} + \Gammao^c_{ab} - h^c_d\; \Gammao^d_{ab} + \order(h^3) \,,\\
  R^a{}_{bcd} &= \Rn^a{}_{bcd} +\Ro^a{}_{bcd} \\
  &\qquad\qquad- \left[h^a_e \; \Ro^{e}{}_{bcd} + \gn{}^{ae} \gn_{fg} \left( \Gammao^g_{ce} \; \Gammao^f_{db} - \Gammao^g_{de} \; \Gammao^f_{cb} \right)\right]+ \order(h^3)\,, \\
  R_{ab} &= \Rn_{ab} + \Ro_{ab} - \left[h^c_d \; \Ro^d{}_{acb} + \gn{}^{ce} \gn_{fg} \left( \Gammao^g_{ce} \;\Gammao^f_{ba} - \Gammao^g_{ce} \; \Gammao^f_{ca} \right)\right]+ \order(h^3)\,, \\
  R &= \Rn +  \Ro + \bigg\{ h^{ac} h_{c}^b\,\Rn_{ab} - h^{ab} \; \Ro_{ab} \\
  &\qquad\qquad\qquad\; - \gn{}^{ab}  \left[h^c_d \; \Ro^d{}_{acb} + \gn{}^{ce} \gn_{fg} \left( \Gammao^g_{ce} \; \Gammao^f_{ba} - \Gammao^g_{be} \; \Gammao^f_{ca} \right)\right]
  \bigg\} + \order(h^3)\,.
\end{split}
\end{equation}
In these expressions the linearized objects are given in terms of $h_{ab}$ by 
\begin{equation}\label{eq:1stOrderQ}
\begin{split}
\Gammao^c_{ab} &\equiv \frac{1}{2} \gn{}^{cd} \left(  \dn_a h_{bd} +  \dn_b h_{ad} -  \dn_d h_{ab} \right) \,, \\
\Ro^a{}_{bcd} &= \dn_{[c}\dn_{d]} h^a_b +  \dn_{[c} \dn_b h^a_{d]} - \dn_{[c} \dn^a h_{d]b}  \,, \\
\Ro_{ab} &= \dn_c \dn_{(a} h^c_{b)} - \frac{1}{2} \Boxn h_{ab} - \frac{1}{2} \dn_a \dn_b h \,, \\
\Ro &= \Ro_{ab} \,\gn{}^{ab} - \Rn^{ab}\, h_{ab}  =  \dn_a \dn_b h^{ab} - \Boxn h - \Rn^{ab} \, h_{ab} \,,
\end{split}
\end{equation}
where $h \equiv h^a_a = \gn{}^{ab} h_{ab}$. 

For the expansion of actions it is also useful to note the following identity:
\begin{equation}
 \sqrt{-g} = \sqrt{-\gn} \left( 1 + \frac{1}{2} h + \frac{1}{8} (h^2 - 2 h^a_b \, h^b_a ) + \order(h^3)\right) \,.
\end{equation}
%

\bibliographystyle{JHEP}
\bibliography{sphereical}

\end{document}